\documentclass[aip,jcp]{revtex4-1}
\usepackage{bm}
\usepackage{graphicx}
\usepackage{dcolumn}
\usepackage{amsmath}

\renewcommand{\vec}[1]{\mathbf{#1}}
\newcommand{\eqnref}[1]{Eq.~\eqref{#1}}
\def\G{{\mathbf G}}
\def\X{{\mathbf X}}
\def\C{{\mathbf C}}
\def\F{{\mathbf F}}
\def\P{{\mathbf P}}
\def\a{\uparrow}
\def\b{\downarrow}

\begin{document} 

\title{Complete spectrum of the infinite-$U$ Hubbard ring using group theory}

\author{Alessandro Soncini}
\email{asoncini@unimelb.edu.au}
\author{Willem Van den Heuvel}
\affiliation{School of Chemistry, The University of Melbourne, VIC 3010, Australia}

\date{\today}

\begin{abstract}
We present a full analytical solution of the multiconfigurational strongly-correlated mixed-valence 
problem corresponding to the $N$-Hubbard ring filled with $N-1$ electrons, and infinite 
on-site repulsion. While the eigenvalues and the eigenstates of the model are known already, 
analytical determination of their degeneracy is presented here for the first time.
The full solution, including degeneracy count, is achieved for each spin configuration 
by mapping the Hubbard model into a set of H\"uckel-annulene problems for 
rings of variable size.  The \emph{number} and \emph{size} of these effective H\"uckel annulenes, 
both crucial to obtain Hubbard states and their degeneracy, are determined 
by solving a well-known combinatorial enumeration problem, the necklace problem for $N-1$ beads 
and two colors, within each subgroup of the $\C_{N-1}$ permutation group.
Symmetry-adapted solution of the necklace enumeration problem is finally achieved by 
means of the \emph{subduction of coset representation} technique [S. Fujita, \emph{Theor. Chem. Acta} {\bf 76}, 247 (1989)], 
which provides a general and elegant strategy to solve the one-hole infinite-$U$ Hubbard problem, including 
degeneracy count, for any ring size. The proposed group theoretical strategy to solve
the infinite-$U$ Hubbard problem for $N-1$ electrons, is easily generalized to the case 
of arbitrary electron count $L$, by analyzing the permutation group $\C_L$ and all its subgroups.
\end{abstract}

\maketitle

\section{Introduction} 
It is a well-established fact that the electronic structure of
systems containing $d$ and $f$ electrons is poorly 
modelled by single-determinant approximations. 
Well-known examples in solid state physics are Mott insulators,\cite{Mott1968} wrongly predicted to be
metals within an independent particle picture. In molecular science, strong electron correlation 
and multiconfigurational electronic states play a central role in the description of the 
rich magnetic behavior of polynuclear inorganic complexes of transition metal and rare earth ions 
with partially filled $d$ and $f$ angular momentum shells, also known as
molecular nanomagnets.\cite{Gatteschi2006}   
Despite the advances of multiconfigurational ab initio methods
such as CASSCF/CASPT2, first principles approaches are to date still too demanding 
to describe complexes involving more than one or two metal ions.  In this scenario, simple models of electron 
correlation can be very helpful, both to provide interpretation of ab initio results, or to tackle
large electronic structure problems.

One widely used multiconfigurational atomistic model of strongly electron-correlated systems 
is the Hubbard model.\cite{EsslerBook2005,Tasaki1998a,Tasaki1998b} In its original formulation it provides 
a description of a set of $L$ active 
electrons occupying $N$ orthogonal orbitals localized on $N$ metal atoms. 
The simplest Hubbard Hamiltonian reads:
\begin{equation}
H = t \sum_{\langle ij \rangle}\sum_{\sigma}^{\uparrow\downarrow} c_{i\sigma}^{\dagger}c_{j\sigma} 
+ U \sum_{i}^N n_{i\uparrow} n_{i\downarrow}
\label{hubham}
\end{equation}
where according to the usual notation $c_{i\sigma}^{\dagger}$ and $c_{i\sigma}$ are the 
creation and annihilation operators for electrons occupying the atomic orbital at site $i$ with spin 
$\sigma$, and $n_{i\sigma}=c^{\dagger}_{i\sigma}c_{i\sigma}$, and the angular parenthesis limits 
summation over nearest-neighbors .
The two fundamental ingredients in the basic Hubbard Hamiltonian are the 
charge transfer term $H_t$ between nearest neighbor sites (first term on the right hand side of \eqnref{hubham}), 
here parametrized by the 
hopping integral $t$, and the on-site Coulomb repulsion term, here parametrized by 
the two-electron repulsion integral $U > 0$. 

\begin{figure}
\vspace{1cm}
\includegraphics[scale=0.52]{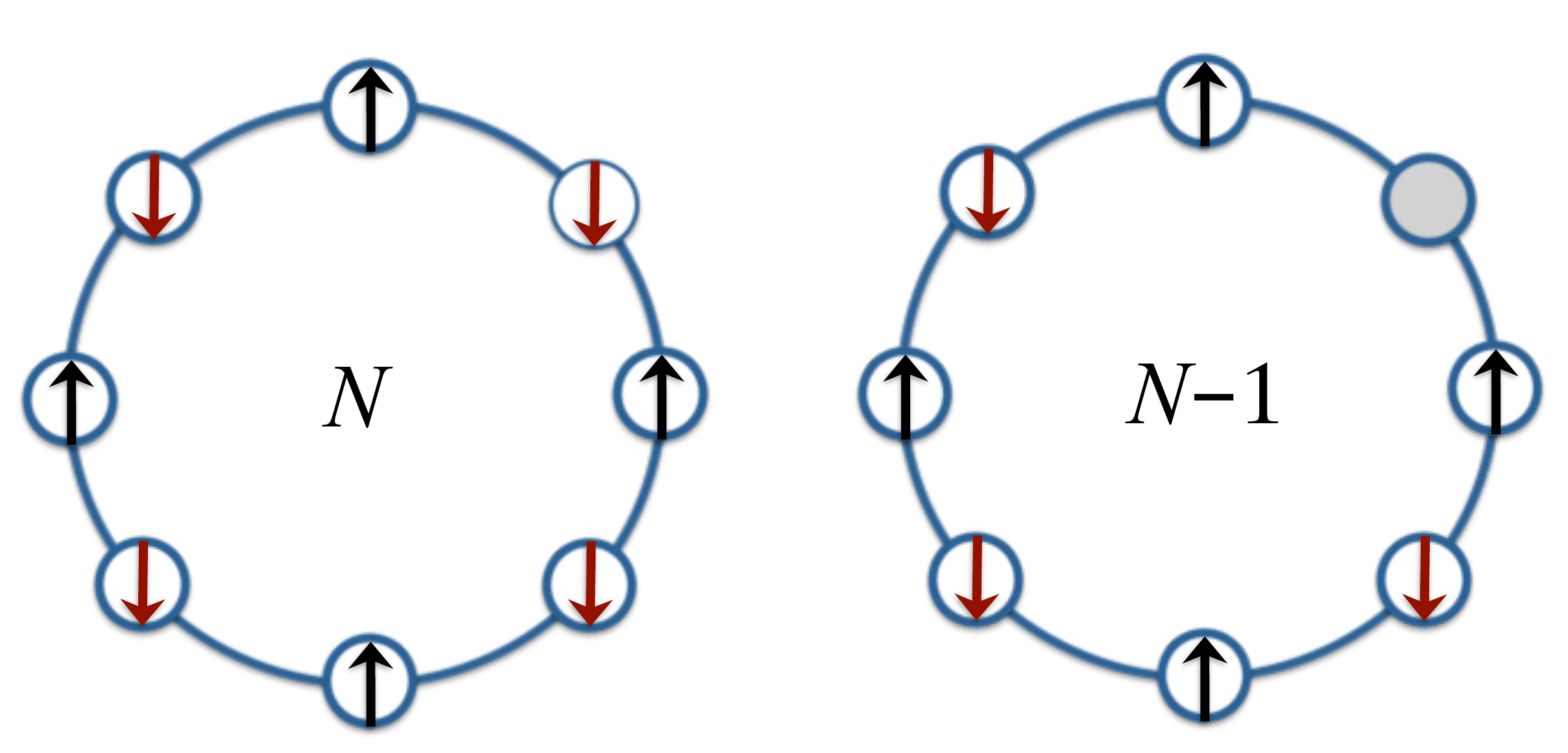}
\caption{Scheme of a Hubbard ring with $N$ metal centers and 
$N$ electrons (left) or $N-1$ electrons (right). The constraint
$U=\infty$ implies that each orbital is occupied at most by one electron, and thus
all electronic configurations describing the multiconfigurational states of the 
$N-1$ Hubbard ring will always involve a single empty orbital (shaded-atom).}
\label{scheme1}
\end{figure}

\subsection{Mixed-valence one-hole Hubbard ring} 
Despite its apparent simplicity, exact solutions to \eqnref{hubham} are known for very few connectivities.  
One well-known case is that of 1-dimensional systems, also known as Hubbard rings, 
representing an important electron correlation model for e.g. molecular 
wheel nanomagnets.  The Hubbard ring problem can be solved exactly either via 
the Bethe ansatz, or in the case of infinite U, also via a particular unitary transformation 
of the basis states.  Here we will be interested in the infinite U case, which represents an approximation 
to the strong coupling limit.  Although the expression for the eigenvalues of the Hubbard
ring with infinite U and arbitrary filling is
known,\cite{Caspers1989,Kotrla1990} to the best of our knowledge the exact degeneracy 
of each solution has never been addressed in the literature.  

Two particular electron-counts are clearly of greater relevance, as these counts
are more likely to represent chemically stable charge-states of molecular metal rings:  
the half-filling electron count ($N$ electrons on $N$ metal centers, see Fig.~\ref{scheme1} left), 
and the half-filling minus one  electron count ($N-1$ electrons on $N$ metal centers,
see Fig.~\ref{scheme1} right).  
Note that the half-filling plus one ($N+1$ electrons) is obtained from 
the $N-1$ case simply by changing the sign of the hopping integral $t$. 

The first case (half-filling) is uninteresting in the limit of infinite U, as then 
all $2^N$ Slater determinants have the same energy, since no hopping process
is permitted by the infinite value of U. In fact the half-filling case in the limit of 
large but finite U can be discussed also within a perturbative approach, where
the hopping part of the Hamiltonian couples the $2^N$ degenerate determinants
arising from single occupation of the orbital, with charge transfer configurations
in which one orbital is doubly occupied and another remains empty. The inclusion 
of the effect of the high-energy charge-transfer configurations to second order
in $t$ leads to the mapping of the Hubbard ring problem for half-filling into the 
Heisenberg ring problem with spin one-half on site.\cite{Tasaki1998b}

More interesting is the second case ($N-1$ electrons) in the limit of infinite U.
This model represents the simplest description of the electron correlation 
problem arising in a mixed-valence metal ring, where one metal contributes
no valence electrons, while all the others contribute one electron.
For instance, singly oxidized (and singly reduced) infinite-$U$ Hubbard rings can be used  
to describe states that are relevant for quantum transport in molecular rings devices in 
the Coulomb-blockade regime,~\cite{BruusFlensberg2004,ElsteTimm2005} as conduction via such rings 
is described by electrodes-induced transitions between the states of the half-filled ring, 
and those of the singly oxidized or singly reduced ring, with the extra electron occupying 
an empty atomic orbital centred at a metal's site.~\cite{SonciniJACS2010,SonciniPRB2010}

In this paper we show that the Hubbard $N$-ring for $N-1$ electron filling can be
solved exactly by mapping it into a set of H\"uckel annulene problems for which the
analytical spectrum is well known once the size of the ring is known. Thus once the 
number $N$ of metal centers in the Hubbard ring is known and the total number of 
spin-up electrons in the ring is fixed, the only problem that remains to be solved  is to 
determine what are the sizes of the associated H\"uckel rings, and how many H\"uckel
rings of a given size are there.  This problem will be solved with the aid of 
group theory.   Finally, we will show that our group theory strategy to count
the repetition of the same effective H\"uckel spectrum in the solution of the $N-1$ Hubbard 
problem can also be applied to count analytical solutions for any electron filling of the ring.

\section{Mapping of the one-hole Hubbard ring problem into a collection of H\"uckel problems}

The Hubbard Hamiltonian~\eqnref{hubham}  can be simplified for a metal ring with $N$ sites as
\begin{equation}\label{hubbardHam}
H=t\sum_{i=1}^N\sum_{\sigma=\uparrow,\downarrow}(c^\dagger_{i+1,\sigma}c_{i,\sigma}+
c^\dagger_{i,\sigma}c_{i+1,\sigma})+U\sum_{i=1}^N
n_{i,\uparrow}n_{i,\downarrow},
\end{equation}
where cyclic boundary conditions are imposed by identifying site $N+1$ with site 1.
The ring is occupied with $L\leq2N$ electrons. The solutions for $L> N$
electrons are obtained easily from the solutions for $2N-L$ electrons by
replacing $t$ with $-t$ everywhere (this is the hole-particle transformation).
We will therefore consider the $L\leq N$ cases only. 

When $U=0$, \eqnref{hubbardHam} trivially reduces to the Hamiltonian of a H\"uckel cycle
with $L$ noninteracting electrons, whose well-known eigenstates consist of
single Slater determinants with energy
\begin{equation}\label{EHuckel}
E=2t\sum_{\lambda}^{\mathrm{occ}} \cos\frac{2\pi \lambda}{N},
\end{equation}
where the sum runs over the $L$ occupied molecular H\"uckel orbitals, labeled
by the quantum number $\lambda$, which can be interpreted as an effective orbital
angular momentum component along the rotational $C_N$ axis of symmetry~\cite{SteinerFowlerJPCA2001,SteinerFowlerChemComm2001}
(and also representing an irreducible representation of the
molecular symmetry group $\C_N$).  The angular momentum $\lambda$ can take the following values:
\begin{align}
\lambda&=0,\pm1,\pm2,\ldots,\pm (N-1)/2  &&\text{for $N$ odd} \label{kLodd}\\
\lambda&=0,\pm1,\pm2,\ldots,\pm (N/2-1),\, N/2  &&\text{for $N$ even}. \label{kLeven}
\end{align}

When $U>0$, the problem becomes multiconfigurational and the solutions are in
general not so easy to find. However, in the limit of strong on-site repulsion 
$U\rightarrow\infty$, the only relevant Slater determinants are those representing
an electronic configuration in which each site-orbital is either empty or singly occupied 
(see Figure~\ref{scheme1} on the right). In this case several useful statements can be 
made about the block-diagonal structure of the Hamiltonian matrix in the basis of
this particular subset of Slater determinants.

Each of these determinants can in fact be specified completely by the row vectors
$\vec{x}=(x_1,x_2,\ldots,x_L)$, listing the occupied sites (in increasing
order), and $\bm{\sigma}=(\sigma_1,\sigma_2,\ldots,\sigma_L$), listing the
corresponding spin values, as follows:
\begin{equation}\label{basis}
|\vec{x},\bm{\sigma}\rangle = c^\dagger_{x_1,\sigma_1}c^\dagger_{x_2,\sigma_2}
\ldots c^\dagger_{x_L,\sigma_L} |0\rangle, \qquad 1\leq x_1<x_2<\ldots <x_L\leq N
\end{equation}

In this work we will focus mainly on the $N-1$-electron count, and for this 
specific case it is possible to classify the one-hole determinant basis states
in terms of the position of the single empty orbital $l$ ($l=1,\dots, N$), and 
the spin configuration $\bm{\sigma}=\{\sigma_i\}_{i \neq l}$ for the $N-1$ sites.
Thus we write the basis of one-hole Slater determinants as:
\begin{equation}
|l,\bm{\sigma}\rangle=(-1)^{l-1}\prod_{\substack{i=1\\i\neq l}}^{N} 
c^\dagger_{i,\sigma_i}|0\rangle.
\label{holebasis}
\end{equation}
We note that this phase choice has the advantage that the matrix elements of $H_t$ 
in this basis are equal either to $-t$ or to zero.\cite{Tasaki1998a}
Each of these states is also characterized by its value of
$M_S=(n_1-n_2)/2$, where $n_1$ ($n_2$) is the
number of spin-up (down) electrons in $\bm{\sigma}$. Both $M_S$ and the total
spin $S$ are conserved quantities.  Within this space we must now diagonalize
the hopping Hamiltonian $H_t$ (first part of \eqnref{hubbardHam}). Under the
action of $H_t$ a spin can hop to a neighboring site only if that site is
empty.  

Since double occupations are never allowed, it follows that in a one-dimensional 
nearest-neighbor connectivity the ordering of a given sequence of spin-up/spin-down
polarizations in $\bm{\sigma}$ will be conserved under the action of $H_t$. We
will also refer to this
$\bm{\sigma}$-ordering as \textit{spin configuration}.  This simple observation has a few crucial consequences:

\begin{itemize}

\item Within a given $M_S$ subspace of Slater determinants, $H_t$ will be block-diagonal in the spin configuration vector $\bm{\sigma}$.

\item The matrix-structure of each $\bm{\sigma}$-block is in fact that of the H\"uckel Hamiltonian matrix for a  $n_\sigma$-annulene, 
with hopping integrals $\beta = -t$.  This can be easily seen by repeated application of $H_t$ to an initial 
one-hole Slater determinant, generating a full closed orbit (H\"uckel annulene) of $n_\sigma$ Slater determinants, where each determinant is only connected by $H_t$ to \emph{two} other determinants: one where the hole is one position back, and the other where the hole is one position forward (see Fig.~\ref{fig2}, illustrating  
the case of 2 electrons in a 3-center Hubbard ring with $M_S=0$, mapped into an $n_\sigma$-annulene with $n_\sigma=6$, i.e. into H\"uckel benzene).   Note that for electron counts different from $N-1$,  $H_t$ still generates a closed orbit of Slater determinants for each given spin configuration $\bm{\sigma}$, although the matrix connectivity of the graph associated to such orbit will not be a simple ring connectivity. 

\item The N-sites Hubbard ring eigenvalues obtained from each block are thus coincident with those of a H\"uckel  annulene problem with $n_\sigma$ sites, and read $\epsilon = -2t\cos \frac{2\pi \lambda}{n_\sigma}$, with $\lambda=0, \pm 1, \pm2, \dots \frac{n_\sigma}{2}$ (if $n_\sigma$ is even), or $\pm\frac{(n_\sigma -1)}{2}$ (if $n_\sigma$ is odd).

\end{itemize}

Hence, the infinite-$U$ Hubbard ring problem is fully diagonalized provided we can  (i) enumerate all the independent H\"uckel rings (i.e. spin configurations $\bm{\sigma}$ ), for every given $M_S$, and (ii) determine the size $n_\sigma$ of each  H\"uckel ring (i.e. the size of the orbit of Slater determinants with same spin configuration $\bm{\sigma}$, generated by repeated application of the hopping Hamiltonian $H_t$).   We show below how this can be simply achieved for small rings, but quickly becomes a non trivial counting problem that needs be approached via the powerful techniques of group theory.

 \subsection{Two electrons in three orbitals: Hubbard 3-ring mapped into H\"uckel benzene}

Let us at first consider the smallest Hubbard ring, with $N=3$ and the non-trivial total spin projection $M_S=0$.
We have here two electrons of opposite spin polarization hopping over three metal-centered orbitals. 
\begin{figure}[ht]
\vspace{1cm}
\includegraphics[scale=0.62]{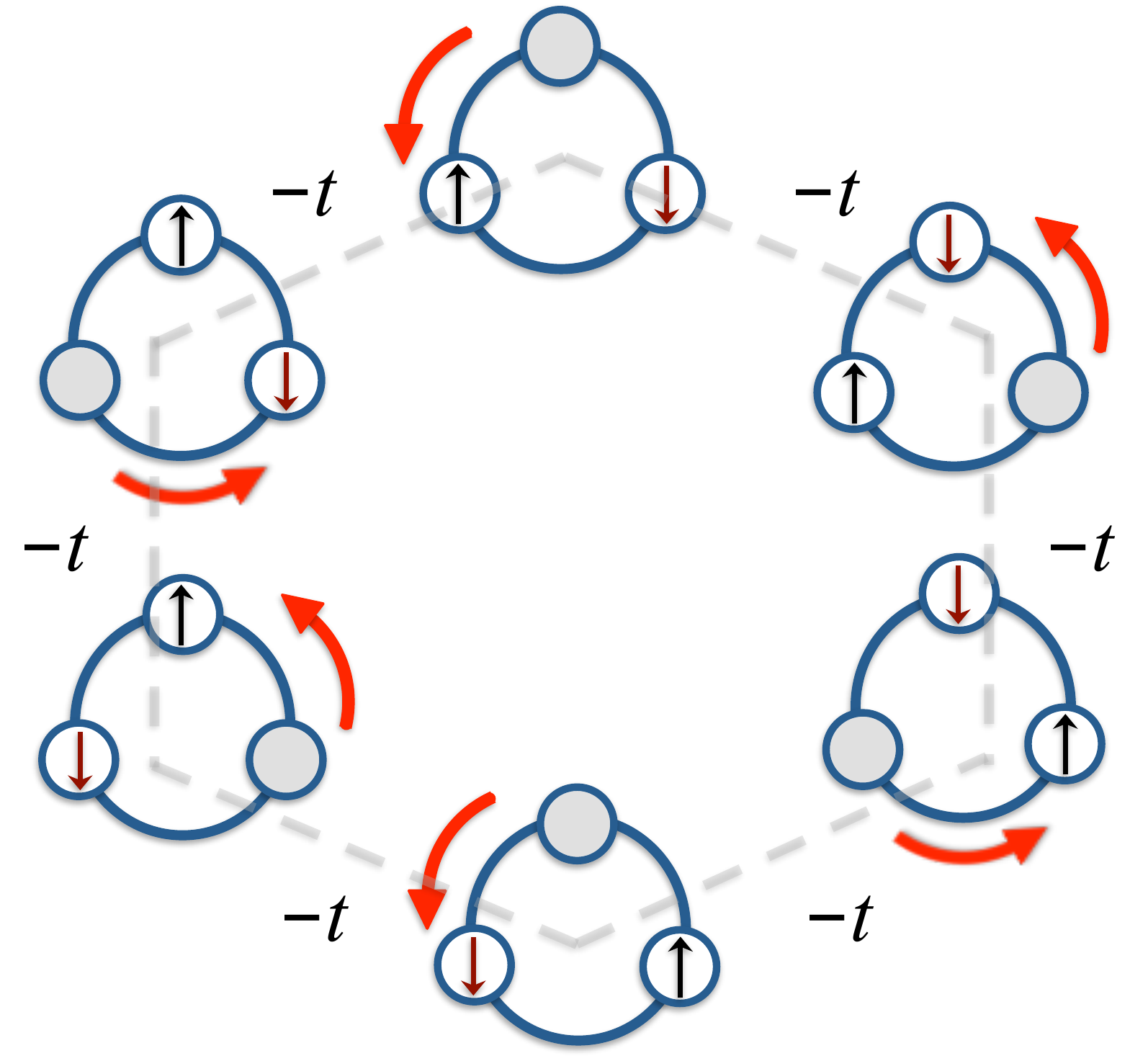}
\caption{The clockwise hopping of two electrons in three orbitals defines an $M_S=0$ 
space of six Slater determinants cyclically connected by hopping integral $-t$.
Note that this is equivalent to the anticlockwise hopping of the empty site around the ring,
although the starting configuration can only be obtained after two turns of the Hubbard ring,
turning the 3-ring into a 6-cycle. Within the determinant basis space, the Hubbard Hamiltonian 
is thus mapped into a H\"uckel benzene, which can be analytically diagonalized (see text).}
\label{fig2}
\end{figure}
For this simple example it is clear that only one orbit of Slater determinants exists. 
It is in fact interesting to note that the hopping of one electron e.g. in a clockwise direction formally 
corresponds to the hopping of the empty orbital in the opposite (anticlockwise) direction, as illustrated in Figure~\ref{fig2}.  

Note also that the empty orbital needs to hop \emph{twice} around the 3-membered ring in order for the hopping 
Hamiltonian $H_t$ to span the whole orbit of Slater determinants, so that the size of this orbit for the spin configuration 
$\bm{\sigma}=(\a, \b)$  is    $n_{\a\b} = 3 \times 2 = 6$.   As anticipated in the previous paragraph, and shown here in Figure~\ref{fig2}, if the 6-dimensional determinant basis is ordered according to consecutive hopping processes, each of the six configurations is connected by $H_t$ only to its two nearest neighbor determinants, so that hopping defines a ring of Slater determinants which has double the size of the Hubbard ring. The resulting block of the infinite-$U$ Hubbard Hamiltonian clearly reads:
\begin{equation}\label{3ringHuckelMat}
H_{\a\b} = \left(
\begin{array}{cccccc}
0      & -t     &  0  &  0    &   0   &  -t  \\
-t     &   0    &  -t  &  0    &   0   &  0  \\
0     &   -t    &   0 &  -t    &   0   &  0  \\
0     &   0    &  -t  &  0    &   -t   &  0  \\
0    &   0    &  0  &  -t    &   0   & -t  \\
-t     &   0    &  0 &  0    &  -t   &  0  
\end{array}
\right)
\end{equation}
which is equivalent to the H\"uckel Hamiltonian for benzene.
Thus  $H_t$ is easily diagonalized within the $M_S=0$ subspace, leading to a spectrum with the six eigenvalues
$\epsilon_\lambda^{\a\b} = -2t \cos(\frac{2\pi \lambda}{6})$, for 
$\lambda=0, \pm 1, \pm2, 3$.  As for the triplet projections $M_S=1$, the matrix representation 
of $H_t$ can be mapped into a H\"uckel $[3]$-annulene, with the three eigenvalues 
$\epsilon_\lambda^{\a\a} = -2t \cos(\frac{2\pi \lambda}{3})$, $\lambda=0, \pm 1$.
Note that if $t>0$,  the ground state is high-spin (triplet), as expected for rings with 
$N=3$ and $N=4$, where Nagaoka's theorem is fulfilled.\cite{Tasaki1998a}

\subsection{Four electrons in five orbitals with $M_S=0$: multiple orbits / H\"uckel annulenes}

\begin{figure}
\vspace{1cm}
\includegraphics[scale=0.62]{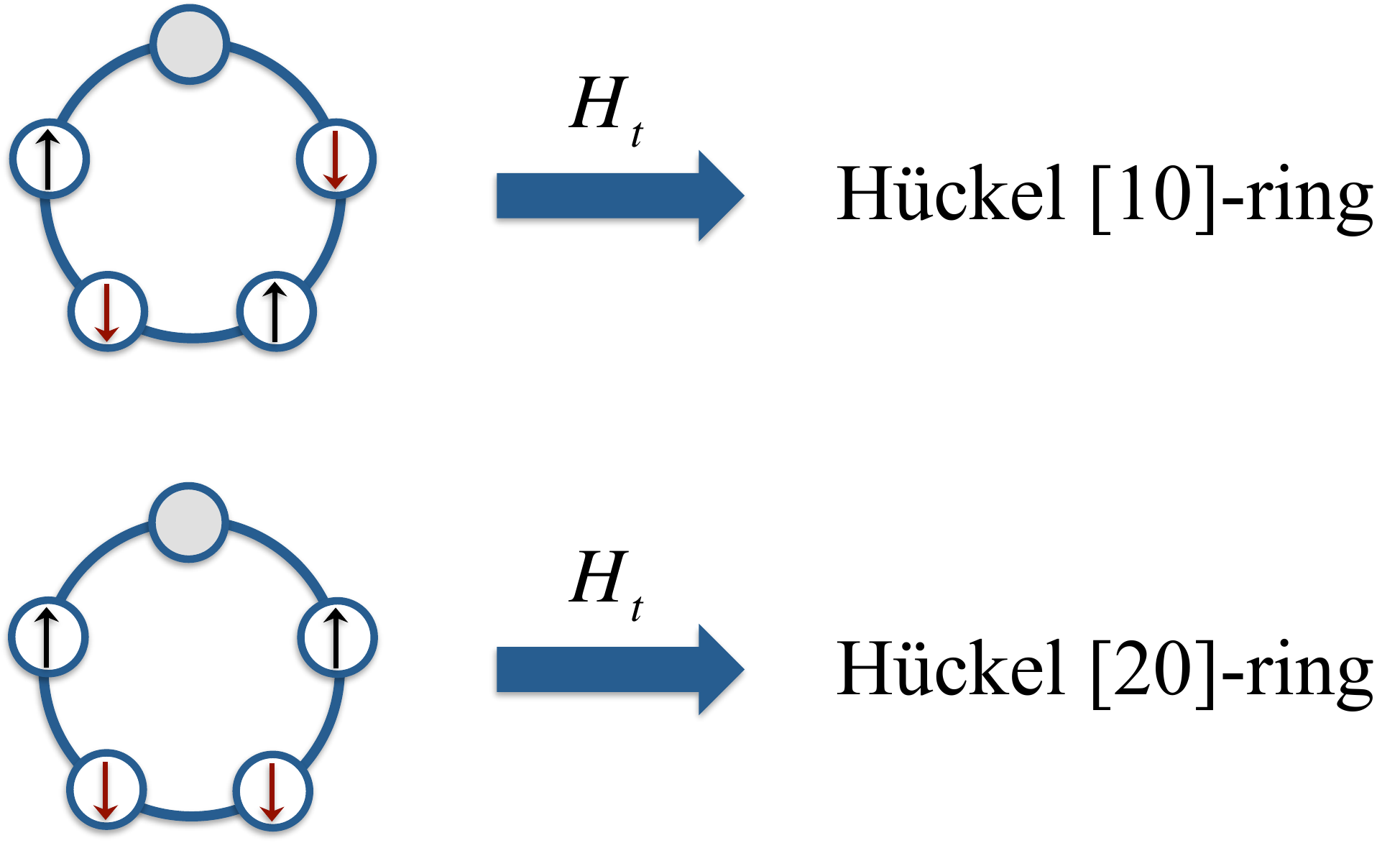}
\caption{The two orbits of spin determinants representing two separate subspaces
within the $M_S=0$ spin projection space, for four electrons in five orbitals.
Note that these two orbits of spin configurations under the action of the 
hopping Hamiltonian $H_t$ generate an adjacency matrix for two 
distinct H\"uckel rings, one with ten vertices, the second with 20 vertices.}
\label{fig3}
\end{figure}

The case of four electrons in a Hubbard ring with five metal centers represents the 
smallest 1D-Hubbard problem for which we encounter multiple orbits/spin configurations 
within a given value of $M_S$. In the case $M_S=0$  we can 
build two families of Slater determinants, one corresponding to an alternating spin configuration
$\bm{\sigma}_1=(\a, \b, \a, \b)$ , the other $\bm{\sigma}_2=(\a, \a, \b, \b)$,  as illustrated in Figure~\ref{fig3}.
It is evident that these two families of determinants cannot be connected via simple hopping process.
Thus the $M_S=0$ subspace is further block-diagonalized into two subspaces, each
subspace corresponding to a different orbit.
 In particular, by repeatedly applying the hopping Hamiltonian $H_t$ to any determinant with
 spin configuration $\bm{\sigma}_1$  (top of Figure~\ref{fig3}), the hole has to hop twice around 
 the 5-Hubbard ring to get back to starting configuration, so that the length of this orbit is
 $n_{\a\b\a\b} = 2 \times 5 = 10$.  The  block $\bm{\sigma}_1$ of the Hubbard Hamiltonian is thus 
 mapped into the eigenvalue problem for the H\"uckel [10]-ring, 
 with spectrum $\epsilon_\lambda^{\a\b\a\b} = -2t \cos(\frac{2\pi \lambda}{10})$, 
 $\lambda=0, \pm 1, \pm 2, \pm 3, \pm4, 5$. On the other hand, it can be seen by direct inspection that 
 the full orbit of determinants corresponding to the spin configuration $\bm{\sigma}_2$  (bottom of Figure~\ref{fig3}) 
 can be generated if the hole hops four times around the 5-Hubbard ring, so that $n_{\a\a\b\b} = 4 \times 5 = 20$.
 The Hubbard block $\bm{\sigma}_2$  is thus equivalent to the Hu\"ckel Hamiltonian for a [20]annulene, with spectrum 
 $\epsilon_\lambda^{\a\b\b\a} = -2t \cos(\frac{2\pi \lambda}{20})$,  $\lambda=0, \pm 1, \pm 2, \dots, \pm9, 10$.

 \section{Mapping the H\"uckel annulenes enumeration problem into a necklace enumeration problem}

 From the previous examples we note that the size of the H\"uckel annulenes associated to the $\bm{\sigma}$-blocks
 is always an integer multiple of  the number of metal centers $N$ in the Hubbard ring, as the hole must always hop in units 
 of $N$-steps to get back to the initial site and close the spin-configuration orbit.  The problem is to find how many times the 
 hole has to hop around the Hubbard ring in order to span the full orbit.  For small 
 Hubbard rings, it is easy enough to work this out by inspection.
 However, the problem becomes increasingly tedious as $N$ becomes larger.  
 
 A systematic strategy to enumerate H\"uckel annulenes and determine their sizes for each given $M_S$ is offered by group theory.  The 
 connection between enumeration of orbits of Slater determinants / H\"uckel annulenes, and group theory, can be readily made 
 by noting that after each single turn of the empty orbital around the Hubbard ring,  
 \emph{the spin configuration $\bm{\sigma}$  undergoes a cyclic permutation} within the remaining $N-1$ occupied sites. 
 
 If all $N-1$ occupied sites have parallel spins ($|M_S|=(N-1)/2$ ), the cyclically permuted spin configuration 
 is indistinguishable from the initial spin configuration, thus a single turn of the empty orbital around the Hubbard ring suffices to generate
 a full orbit of Slater determinants, and the associated H\"uckel ring has the same size as the Hubbard ring (i.e.  $n_\sigma = N$).  
 This implies that the spectrum of the $N$-Hubbard ring with $N-1$ electrons for $|M_S|=(N-1)/2$ corresponds 
 to the H\"uckel spectrum of an $[N]$-annulene with resonance integral $\beta = -t$. 
 The eigenvalues are therefore 
 $\epsilon_\lambda=-2t \cos\left(\frac{2\pi \lambda}{N}\right)$, with $\lambda= 0, \pm1, \dots, N/2$ if $N$ is even, or $\pm (N-1)/2$ if N is odd. 
 Beside the pure spin $\pm M_S$ double degeneracy, these states present additional orbital double-degeneracies associated to 
 the axial orbital angular momentum quantum number $\pm \lambda$, as it is found in common H\"uckel $[N]$-annulenes. 
 
 For $|M_S| <  (N-1)/2$, the cyclically permuted spin configuration $\bm{\sigma}$ generated 
 by $N$ hopping processes is not equivalent to the initial spin configuration. The effect of $N$-hopping processes on a one-hole determinant 
 is thus equivalent to the action of the cyclic permutation $\hat{C}_{N-1}$ (generator of the permutation group $\C_{N-1}$) on 
 a two-color necklace with $N-1$ beads, where a given bead has color X (Y) if the corresponding occupied site in the Hubbard ring 
 has spin up (down).  The two-colored necklace is in fact a representation of the spin-ordered configuration $\bm{\sigma}$ under scrutiny, 
 where the ratio between the number of beads with different color is fixed by the value of $M_S$. 
  
  Crucially, the problem of enumerating $\bm{\sigma}$ spin configurations for a given $M_S$ that are not connected by hopping processes
  (i.e.\ enumerating H\"uckel annulenes), is now mapped into the well-known combinatorial problem of enumerating symmetry-unique 
  (i.e.\ not related by cyclic permutations)  necklaces with $N-1$ beads of two colors, with a fixed ratio between beads of different colors.    
   Furthermore,  grouping together all necklaces of like symmetry, that is all distinguishable necklaces that can be rotated into each other  
  by repeated application of a cyclic permutation $\hat{C}_{N-1}$, we obtain orbits of two-color necklaces with size $\omega_\sigma$. 
  The length of each H\"uckel annulene associated with a fixed $M_S$ Hubbard problem can now be found by determining the length 
  $\omega_\sigma$ of the associated necklace-orbit generated by the action of the cyclic permutation group $\C_{N-1}$ on a representative
  necklace configuration.  Once the length of each necklace orbit has been determined, the size of the associated 
  H\"uckel ring $n_\sigma$, thus the corresponding set of Hubbard eigenvalues, is easily determined as:
  \begin{equation}
  \label{HuckelOrbit}
  n_\sigma = N \times \omega_\sigma
  \end{equation}
  \begin{figure}[ht]
  \vspace{1cm}
  \includegraphics[scale=0.62]{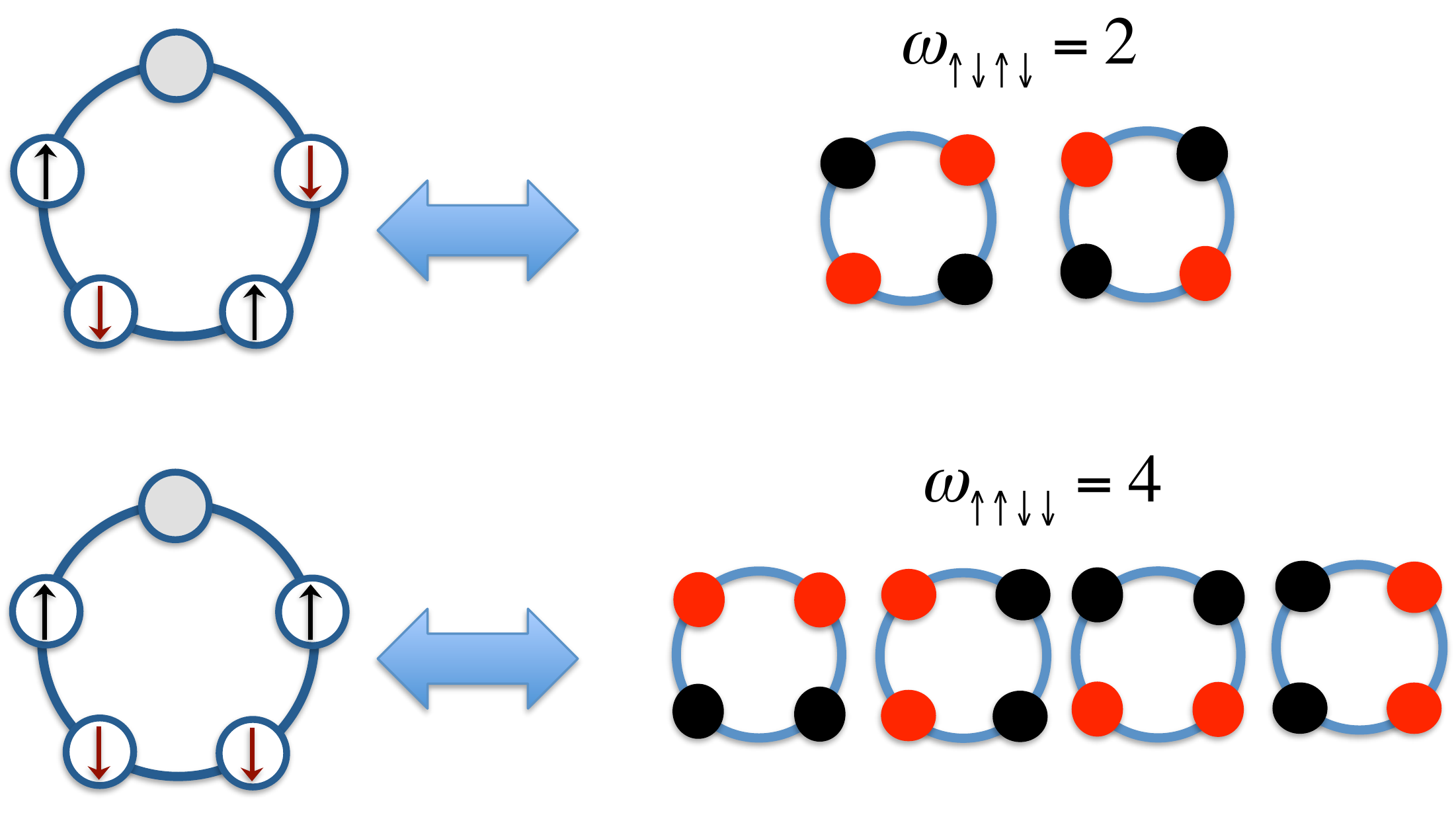}
  \caption{The two 4-beaded necklace orbits associated with the two spin
  configurations spanning the $M_S=0$ spin projection space, for the infinite-$U$ Hubbard problem of four active electrons in five orbitals. The necklace orbit-length is reported as $\omega_\sigma$. To obtain the full
  Slater determinant orbit length (hence the length of the associated H\"uckel rings) the necklace orbit length $\omega_\sigma$ must be
  multiplied by the number of metal centers (i.e.\ by five in this case), as detailed in \eqnref{HuckelOrbit}.}
  \label{fig4}
  \end{figure}
  To illustrate the mapping of the H\"uckel annulenes enumeration problem, into a necklace enumeration problem,
  let us consider the two Slater determinant orbits found in the previous section for the case of four electrons in five active orbitals 
  (see also Figure~\ref{fig3}).
  The mapping process is shown in Figure~\ref{fig4}, where beads of color X and Y are represented by black and red beads.  
  Here we have $N-1 = 4$, so we start off with a 4-beaded necklace which has full permutation symmetry $\C_4$ if all beads 
  have the same color ($|M_S|=2$).   For the $M_S=0$ space, the two spin configurations $\bm{\sigma}_1=(\a, \b, \a, \b)$ 
  and $\bm{\sigma}_2=(\a, \a, \b, \b)$ identified in the previous paragraph can now be mapped into two symmetry-unique 
  necklace configurations, with two black beads, and two red beads. In fact, decoration of the $\C_4$ necklace backbone 
  with beads of two different colors can only lead to necklaces whose symmetry is described by a subgroup of $\C_4$. The 
  group $\C_4$ has three subgroups :  $\C_4$, $\C_2$ and  $\C_1$ (i.e. no symmetry).
  By inspection, it is clear that the necklace associated to $\bm{\sigma}_1=(\a, \b, \a, \b)$ (top of Figure~\ref{fig4}) has 
  permutation symmetry $\C_2$, while the necklace associated to $\bm{\sigma}_2=(\a, \a, \b, \b)$ has symmetry $\C_1$.
  
  The size of the necklace orbits can be easily found by inspection in this case. If we consider the four symmetry operations of
  the group $\C_4 = \left\{ \hat{E}, \hat{C}_4, \hat{C}_4^2\equiv\hat{C}_2, \hat{C}_4^3 \right\}$, by definition the necklace with 
  symmetry $\C_2$ will be invariant with respect to the action of identity and $\hat{C}_4^2\equiv\hat{C}_2$, and will only be rotated
  into a distinguishable configuration under the action of $\hat{C}_4$. The
  orbit is consequently composed of two configurations only
  ($\omega_{\sigma_1} = 2$, see top of Figure~\ref{fig4}). On the other hand, the necklace with symmetry $\C_1$ will be rotated into 
  four symmetry-related but distinguishable necklaces by the action of $\C_4$, thus generating an orbit of size $\omega_{\sigma_2} = 4$ 
  (see bottom of Figure~\ref{fig4}). According to \eqnref{HuckelOrbit}, the size of the corresponding H\"uckel annulenes can then 
  be found as $n_{\sigma_1} = 5 \times \omega_{\sigma_1} = 10$, and $n_{\sigma_2} = 5 \times \omega_{\sigma_2} = 20$.
  
  This reasoning can be made more rigorous within group theory, by exploring the relationship between groups, subgroups and orbits. 
  It is in fact well known that in a structure with a given symmetry group (e.g. a molecule), 
  orbits of symmetry-related points (e.g. atoms, atomic orbitals, bonds, etc.) can be fully characterised in terms of those 
  subgroups describing the site or local symmetry of these points.~\cite{Burnside1911,Fujita1989} 
  Group/subgroup relationships describing orbits in molecular graphs have been used to characterise 
  fundamental chemical and physical properties of molecules.~\cite{FowlerQuinn1986,Fujita1989,CeulemansFowler1991}
  
  More specifically, given a high-symmetry structure described by the group $\G$,
  there is a well-defined link between (i) the symmetry descent from $\G$ to one 
  of its subgroups $\G_k$ describing symmetry-lowering of the structure upon decoration ($\G_k$ plays the role of a site-symmetry), 
  and (ii) the size of the orbits generated by the action of the higher symmetry group $\G$ on the lower-symmetry decorated 
  structures.  In particular, the size of the orbit spanned by the $\G_k$-symmetry decorated structures, is simply the ratio between the 
  number of elements in the higher group generating the orbit (order $|\G|$ of the higher group), and the order $|\G_k|$ of the subgroup. 
  In brief,  $\omega_k = |\G|/|\G_k|$.  In this case, since $\C_4$ has order 4, $\C_2$ has order 2, and $\C_1$ has 
  order 1, it follows that $\omega_{\sigma_1} = 4/2 = 2$, while $\omega_{\sigma_2} = 4/1 = 4$.  
  This useful group/subgroup relationship is analyzed in depth in the following paragraph,  and used to devise 
  a general group theoretical strategy for the enumeration of the spin-configuration necklace 
  orbits and determination of their size, providing the  full spectrum of the infinite-$U$ Hubbard ring, for any size of the 
  Hubbard ring, and any value of $M_S$.

\section{Solution of the necklace enumeration problem}\label{sec:necklace}

In the previous paragraph we have established that each spin configuration $\bm{\sigma}$ 
within a given spin projection $M_S$ is mapped into necklace configurations consisting of $L$ sites 
(so far we have only considered the case $L=N-1$) decorated with beads of at most
two different colors, X and Y (spin-up and spin-down).  The decorated necklaces
can be considered as derivatives of a skeleton with given symmetry $\G = \C_L$,
whose sites are collected in the domain $\Delta = \{1, 2 \dots L\}$. Each
configuration can thus be associated with a symmetry-lowering function $f:
\Delta\rightarrow\X$, mapping each element of the domain $\Delta$ with full
symmetry $\G$, to one of the two elements of the co-domain $\X = \{X, Y\}$. Let
us name $f^{n_1,n_2}_j$ the $j$-th function covering the $L$-beads necklace
with $n_1$ beads of color $X$, and $n_2$ beads of color $Y$, where $n_1+n_2 =
L$, and $M_S = \frac{1}{2}(n_1-n_2)$.  A general skeleton of symmetry $\G$ can
have several sets of symmetry equivalent points, also named {\it orbits}.
In particular, the necklace configurations resulting from the two-color decoration
process, for each value of $M_S$, form a set $\F_{M_S} =
\{f_1^{n_1,n_2},f_2^{n_1,n_2},\dots,f_{|\F_{M_S}|}^{n_1,n_2}\}$, which can have
several sets of equivalent `points' or necklace-orbits. 
The total number of inequivalent orbits for a given weight
$X^{n_1}Y^{n_2}$ can be determined using the
P{\'o}lya-Redfield theorem, by reading out the coefficient of $X^{n_1}Y^{n_2}$
from a so-called Cycle Index computed using appropriate figure inventories and the
cycle-structure of the permutation representation of $\C_L$ on $\Delta$.\cite{Harary1973}
However, such approach will not help us here to determine the length of each orbit,
which is the key piece of information to determine the H\"uckel annulene lengths, thus 
their energies.  The problem can instead be solved by counting the orbits of weight
$X^{n_1}Y^{n_2}$  \emph{for each allowed symmetry} describing  
the orbits of necklace configurations in $\F_{M_S}$.  This can be done by partitioning the 
orbit-counting for a given weight, within each subgroup of $\C_L$. 

The method to achieve this symmetry-classified orbit-counting has been proposed 
by Fujita.\cite{Fujita1989} In the next two sub-paragraphs we will introduce a rigorous 
classification of orbits according to group/subgroup relationships, briefly sketch 
the basic features of Fujita's strategy to count orbits within separate symmetry subgroups, 
and apply it to the present case.

\subsection{Rigorous group-theoretical classification of orbits: the coset representations}
A point group $\G$ of order $L$ can be characterized by a non-redundant set of
$s$ subgroups $\{\G_1, \G_2,\dots \G_s \}$, each of which, in turn, gives rise
to a (right) coset-decomposition of the group $\G$:
\[
\G = \G_k g_0 + \G_k g_1 + \dots + \G_k g_{m-1}
\]
where, if $n_k$ is the order of the subgroup $\G_k$, $g_j$ are the $m=L/n_k$
representatives (or transversals) of the $m$ cosets associated to the subgroup
$\G_k$, with $g_0$ the identity operator.  Thus each set of cosets $\G/\G_k =
\{\G_k,\G_k g_1, \dots \G_k g_{m-1}\} (k=1,\dots,s)$, under the action of the
group $\G$, defines a permutation representation $\G(/\G_k) = \{p_g, \forall
g\in\G\}$, where each operator $g\in\G$ is associated to the permutation $p_g$
in the following manner:
\begin{equation}\label{permutation}
p_g = \G(/\G_k)_g = \left(
\begin{array}{cccc}
\G_k g_0 & \G_k g_1 & \dots & \G_k g_{m-1}\\
\G_k g_0 g & \G_k g_1 g & \dots & \G_k g_{m-1} g
\end{array}
\right)
\end{equation}

When $\G_k$ is the identity group $\C_1$, the coset representation $\G(/\C_1)$ is also known as the {\it regular representation}.
Two facts about coset representations (CRs) are well known. First, CRs are all transitive representations
(i.e.\ for any two cosets there exists a $g\in\G$ that connect them). Second, suppose the action of a group $\G$ on a 
set $\Delta$ results in a partition of $\Delta$ into orbits. Then each transitive permutation representation originating
from the action of $\G$ on a particular orbit is equivalent to one of the coset representations $\G(/\G_j)$, and the 
subgroup $\G_j$ describes the `local' or `site' symmetry of each member of the orbit.

Thus any permutation representation $\P_G$ of the group $\G$ resulting from the action of $\G$ onto a domain $\Delta$ 
composed
of multiple orbits, can be `reduced' to a sum of coset representations (see
Theorem 2 in Ref.~\onlinecite{Fujita1989}):
\begin{equation}\label{crdecomposition}
\P_G = \sum_{i=1}^{s} \alpha_i \G(/\G_i),
\end{equation}
where the $\alpha_i$ are the multiplicities describing how many times the orbit $\Delta_i$, described by the 
coset representation $\G(/\G_i)$, appears in the decomposition of the domain $\Delta$.
It can be shown that the multiplicities $\alpha_i$ can be determined by solving the following system of linear equations:
\begin{equation}\label{reductio1}
\mu_j = \sum_{i=1}^{s} \alpha_i m_{ij}, \;\; j=1, 2,\dots s,
\end{equation}
where $\mu_j$ represent the number of points in $\Delta$ that remain fixed under the action of all 
operations of the subgroup $\G_j$
(also known as the {\it mark} of $\G_j$ in $\P_G$), and  $m_{ij}$ is the mark
(number of fixed points) of $\G_j$ in $\G(/\G_i)$.\cite{Burnside1911} Note
that, whereas $\mu_j$ depends on the specific choice of $\Delta$ for the
problem at hand, the marks $m_{ij}$ are solely dependent on the fundamental
structure of the group $\G$ and its relation to its subgroups, thus can
be computed once and for all (tables of marks are reminiscent of character
tables, and the determination of the $\alpha_i$ is reminiscent of a reduction
to irreducible representations).

\subsection{Orbits of two-color necklaces}

Given the set of configurations/necklaces with a certain weight $\theta=(n_1,n_2)$
(partition of $L=n_1+n_2$), $\F_{M_S} =
\{f_1^{\theta},f_2^{\theta},\dots,f_{|\F_{M_S}|}^{\theta}\}$, we want to (i)
consider this set as a new domain of `points' $\Delta'$,  (ii) generate a
permutation representation $\Pi_G^{\theta}$ of $\G \equiv \C_L$ acting on the domain
$\Delta'$ (iii) decompose the permutation representation $\Pi_G^{\theta}$ into
coset representations multiplied by multiplicities $A_{\theta i}$, and (iv)
finally, find a reduction formula like~\eqnref{reductio1} providing a strategy
to compute the multiplicities $A_{\theta i}$ from known information concerning
the structure of the group $\G$, such as the table of marks. Note that the
multiplicities $A_{\theta i}$ are in fact the solutions to our problem, as they
provide the number of orbits of configurations with given spin $M_S$ (i.e.,
given weight $\theta$), for each subgroup $\G_k$ of the parent group $\G$,
and thus the length of each orbit as $\omega_{k} = |\G|/|\G_k|$.

We start off by defining the permutation representation $\Pi_G^{\theta}$ of $\G$ acting on the domain 
of configurations $\F_{M_S}$.  Given a domain $\Delta$, a co-domain $\X$ and the functions 
$f^{\theta}_k:\Delta \rightarrow \X$, with $f^{\theta}_k\in\F_{M_S}$, 
consider a permutation $p_g \in \P_G$. We can define a permutation $\pi_g\in\Pi_G$ as:
\[
\pi_g = \left(
\begin{array}{cccc}
f^{\theta}_1(\delta) & f^{\theta}_2(\delta)  & \dots & f^{\theta}_{|\F_{M_S}|}(\delta)\\
f^{\theta}_1(p_g(\delta)) & f^{\theta}_2(p_g(\delta))  & \dots & f^{\theta}_{|\F_{M_S}|}(p_g(\delta))
\end{array}
\right)
\]
Straightforward application of~\eqnref{crdecomposition} allows us to decompose the permutation 
representation $\Pi_G^{\theta}$ on $\F_{M_S}$ into coset representations
(orbits), according to: 
\begin{equation}\label{crdecomposition2}
\Pi_G^{\theta} = \sum_{i=1}^{s} A_{\theta i}\, \G(/\G_i),
\end{equation}
and to write a reduction formula which allows the calculation of the multiplicities $A_{\theta i}$:
\begin{equation}\label{reductio2}
\rho_{\theta\! j} = \sum_{i=1}^{s} A_{\theta i} m_{ij},
\end{equation}
where the marks $\rho_{\theta j}$ are the number of fixed configurations in
$\Pi_G^{\theta}$ under the action of the subgroup $\G_j$.  Although
\eqnref{reductio2} allows in principle the calculation of the
symmetry-partitioned orbit multiplicities $A_{\theta i}$, as Fujita points out in his work\cite{Fujita1989},
due to the abstract nature of the configurations $f^{\theta}_k\in\F_{M_S}$ it
is in general not straightforward to compute the marks $\rho_{\theta\! j}$.  
A powerful strategy to obtain the $\rho_{\theta\! j}$ is based on the subduction of
the coset representations of $\G$ under the subgroups $\G_i$ in combination
with a P\'olya-Redfield type counting methodology. This strategy, which is due to
Fujita\cite{Fujita1989}, is presented in Appendix B. 

The problem we are currently interested in, the two-colored necklace of length
$L$, is sufficiently simple to allow a direct computation of the
$\rho_{\theta\!  j}$. We recall that $\G$ is in this case the cyclic group
$\C_L$, whose subgroups are the cyclic groups $\C_j$, $\forall j\mid L$. (The
notation $j\mid L$ means ``$j$ is a divisor of $L$''.) The domain
$\Delta=\{1,2,\ldots,L\}$ consists of the sites of the necklace and transforms
as one orbit (corresponding to the regular representation of $\C_L$). A
function $f_k^\theta$, $\theta=(n_1,n_2)$, colors $n_1$ sites black and $n_2$
sites red.  The question is now, for a given $\theta$, how many such
colorings $f_k^\theta$ are invariant under the action of $\C_j$. The action of
$\C_j$ on $\Delta$ divides $\Delta$ in $L/\!j$ suborbits of length $j$. For a
coloring to be invariant under $\C_j$, all sites of the same suborbit must have
the same color. Hence $j$ must be a divisor of $n_1$. Then $n_1/\!j$ of $L/\!j$
suborbits must be colored black and the number of ways to do this is the
sought-after $\rho_{\theta\! j}$:
\begin{equation}\label{rho}
\rho_{\theta\! j}=\begin{cases}
  \dbinom{L/\!j}{n_1/\!j}& \text{if $j\mid n_1$},\\
  0& \text{otherwise}.
 \end{cases}
\end{equation}

To find $A_{\theta i}$, which gives the number of inequivalent colored
necklaces of weight $\theta$ and symmetry $\C_i$, we invert
\eqnref{reductio2}:
\begin{equation}\label{A}
A_{\theta i}=\sum_{j\mid L} \rho_{\theta\! j} \overline{m}_{ji}
\end{equation}
Note that we are adopting a different labeling here: $j$ is the
order of the subgroup rather than a generic index as in \eqnref{reductio2}.
This choice is more convenient in working with cyclic groups. The marks are
computed in Appendix A and given by \eqnref{marktable}, which we report here for convenience: 
\begin{equation}\label{eq:marks}
m_{ij}=\begin{cases}
  L/i & \text{if $j\mid i$}, \\
  0 & \text{otherwise}.
 \end{cases}
\end{equation}
The inverse matrix is defined by $\sum_{j\mid
L}m_{ij}\overline{m}_{jk}=\delta_{ik}$, which can be rewritten using
\eqnref{eq:marks} as $\sum_{j\mid i}\overline{m}_{jk}=(i/L)\delta_{ik}$.
We now apply the M\"obius inversion formula \cite{Hardy2008} to this equation,
which gives $\overline{m}_{ik}=\sum_{j\mid i}(j/L)\mu(i/j)\delta_{jk}$, or
\begin{equation}\label{inversem}
\overline{m}_{ji}=\begin{cases}
  \mu\Bigl(\dfrac{j}{i}\Bigr) \dfrac{i}{L} & \text{if $i\mid j$},\\
  0 & \text{otherwise},
 \end{cases}
\end{equation}
where $\mu(d)$ is the M\"obius function ($d$ is an integer).\cite{Hardy2008} 
Substituting
\eqref{rho} and \eqref{inversem} in \eqref{A} yields 
\begin{equation}\label{Amulti}
A_{\theta i} = \frac{i}{L} \sum_{\substack{j\mid L\\j\mid
n_1}}\binom{L/\!j}{n_1/\!j} \tilde{\mu}\Bigl(\frac{j}{i}\Bigr),
\end{equation}
where, for convenience of notation, we have extended the M\"obius function over
the domain of rational numbers: $\tilde{\mu}(x)=\mu(x)$ if $x$ is an integer and 0 otherwise.

For the case of the infinite-$U$ $N$-Hubbard ring with $L=N-1$ electrons, we will have for each value of $M_S$ 
($n_1(M_S) = M_S + \frac{L}{2}$), and for each subgroup $\C_k \subset \C_{N-1}$ (i.e. for each $k\mid (N-1)$), 
exactly $A_{\theta k}$ copies of a H\"uckel $\left[N(N-1)/k\right]$-annulene spectrum given by:
\begin{equation}
\label{fullhubbardNm1}
\epsilon_{\lambda,k}^{M_S}=-2t\cos\left[ \frac{2\pi k \lambda}{N (N-1)}  \right] 
\end{equation}
where the axial orbital angular momentum quantum number $\lambda$ : 
\[
\lambda = 0, \pm1, \pm 2, \dots, \frac{N(N-1)}{2 k}
\]
if $\left[N(N-1)/k\right]$ is even, while
\[
\lambda = 0, \pm1, \pm 2, \dots, \pm\frac{1}{2}\times\left[\frac{N(N-1)}{ k}-1\right]
\]
 if $\left[N(N-1)/k\right]$ is odd, with multiplicity:
 \begin{equation}
 \label{multifullhubbardNm1}
A_{\theta k} = \frac{k}{N-1} \sum_{\substack{j\mid N-1\\j\mid
n_1(M_S)}}\binom{(N-1)/\!j}{n_1(M_S)/\!j} \tilde{\mu}\Bigl(\frac{j}{k}\Bigr),
\end{equation}

\subsection{Examples}
In this section we will illustrate the use of~\eqnref{rho}--\eqnref{multifullhubbardNm1} to analytically 
determine the full spectrum of one-hole infinite-$U$ Hubbard rings for a few values of $N$.

Let us consider as an example the case of the 6-beads necklace, corresponding
to a Hubbard ring with 7 metal centers and 6 electrons.
The relevant cyclic group is thus $\C_6$, with the four subgroups
$\{\C_1, \C_2, \C_3, \C_6 \}$. 
The possible configurations
are $X^6 Y^0 (M_S=3)$, $X^5 Y^1 (M_S=2)$, $X^4 Y^2 (M_S=1)$,
and $X^3 Y^3 (M_S=0)$.  The inverse mark table for $\C_6$ is (table of marks is
computed in Appendix):
\begin{equation}\label{invmarks6}
\overline{\mathbf{m}} = \left( \begin{array}{rrrr}
\frac{1}{6}  &  0  &  0  &  0\\
-\frac{1}{6} &  \frac{1}{3} &  0  &  0\\
-\frac{1}{6}  &  0  &  \frac{1}{2}  &  0\\
\frac{1}{6}  &  -\frac{1}{3}  &  -\frac{1}{2}  &  1
\end{array}\right)
\end{equation}
and the $\boldsymbol\rho_{\theta}$ vectors are:
\begin{equation}\label{fixed-configuration6}
\begin{array}{rrrrrrrl}
\boldsymbol\rho_{6,0} & = & ( & 1 & 1 & 1 & 1 & ) \\ 
\boldsymbol\rho_{5,1} & = & ( & 6 & 0 & 0 & 0 & ) \\ 
\boldsymbol\rho_{4,2} & = & ( & 15& 3 & 0 & 0 & ) \\ 
\boldsymbol\rho_{3,3} & = & ( & 20& 0 & 2 & 0 & ).
\end{array}
\end{equation}
By multiplying the fixed-configuration vectors~\ref{fixed-configuration6} times the
inverse marks table we obtain at once all the orbits classified 
by subgroup, with the order $\{\C_1, \C_2, \C_3, \C_6 \}$, where the orbit
size is $\{6,3,2,1\}$, as:
\begin{equation}\label{solutions6}
\begin{array}{rrrrrrrl}
\mathbf{A}_{6,0} & = & ( & 0 & 0 & 0 & 1 & ) \\ 
\mathbf{A}_{5,1} & = & ( & 1 & 0 & 0 & 0 & ) \\ 
\mathbf{A}_{4,2} & = & ( & 2 & 1 & 0 & 0 & ) \\ 
\mathbf{A}_{3,3} & = & ( & 3 & 0 & 1 & 0 & ).
\end{array}
\end{equation}

The length of the orbits of Slater determinants for the 7-membered Hubbard ring with
6 electrons are subsequently obtained by multiplying each orbit length by 7, therefore 
leading in this case to: 1 H\"uckel cycle of length 7 ($M_S=3$), 1 cycle of length 42 ($M_S=2$),
2 cycles of length 42, and 1 cycle of length 21 ($M_S=1$), and 3 cycles with length 42 and
1 cycle with length 14 ($M_S=0$), thus the energies in units of $-2 t$ (in parenthesis beside $M_S$
we give the degeneracy of each state $\lambda$):
\begin{equation*}
\begin{array}{lccl}
M_S=3 (1)&:& \epsilon_{\lambda} = \cos\left(\frac{2\pi\lambda}{7} \right) & \lambda=0, \pm1,\pm 2,\pm 3  \\
M_S=2 (1)&:& \epsilon_{\lambda} = \cos\left(\frac{2\pi\lambda}{42}\right) & \lambda=0, \pm1,\dots, \pm 20, 21 \\
M_S=1 (2)&:& \epsilon_{\lambda} = \cos\left(\frac{2\pi\lambda}{42}\right) & \lambda=0, \pm1,\dots, \pm 20, 21 \\
M_S=1 (1)&:& \epsilon_{\lambda} = \cos\left(\frac{2\pi\lambda}{21}\right) & \lambda=0, \pm1,\dots, \pm 9, \pm 10 \\
M_S=0 (3)&:& \epsilon_{\lambda} = \cos\left(\frac{2\pi\lambda}{42}\right) & \lambda=0, \pm1,\dots, \pm 20, 21 \\
M_S=0 (1)&:& \epsilon_{\lambda} = \cos\left(\frac{2\pi\lambda}{14}\right) & \lambda=0, \dots,\pm 6, 7 
\end{array}
\end{equation*}
Note that $\lambda$ is an effective angular momentum in the configuration space which gives rise to a real 
energy degeneracy.

Another example is given here consisting of a 13-center Hubbard ring with 12 electrons.
The parent symmetry of the necklace problem is $\C_{12}$, with 6 subgroups $\{\C_1,\C_2,\C_3,\C_4,\C_6,\C_{12}\}$.
The possible configurations $\theta$
are $X^{12} Y^0 (M_S=6)$, $X^{11} Y^1 (M_S=5)$, $X^{10} Y^2 (M_S=4)$, 
$X^9 Y^3 (M_S=3)$, $X^8 Y^4 (M_S=2)$, $X^7 Y^5 (M_S=1)$, and $X^6 Y^6 (M_S=0)$.  
The inverse mark table for $\C_{12}$ is (see table of marks in Appendix A):
\begin{equation}\label{invmarks12}
\overline{\mathbf{m}} = \left( \begin{array}{rrrrrr}
\frac{1}{12}  &  0  &  0  &  0 & 0 & 0\\
-\frac{1}{12} &  \frac{1}{6} &  0  &  0 & 0 & 0\\
-\frac{1}{12}  &  0  &  \frac{1}{4}  &  0 & 0 & 0\\
      0       &  -\frac{1}{6}  &  0  &  \frac{1}{3} & 0 & 0\\
\frac{1}{12}  &  -\frac{1}{6}  &  -\frac{1}{4}  &  0 & \frac{1}{2} & 0\\
      0        &  \frac{1}{6}  & 0  &  -\frac{1}{3}  & -\frac{1}{2}  &  1 
\end{array}\right)
\end{equation}
and the $\boldsymbol\rho_{\theta}$ vectors of fixed configurations
under the action of the subgroups $\{\C_1,\C_2,\C_3,\C_4,\C_6,\C_{12}\}$ are:
\begin{equation*}
\begin{array}{lrrrrrrrrl}
\boldsymbol\rho_{12,0} & = & ( & 1 & 1 & 1 & 1 & 1 & 1 &) \\ 
\boldsymbol\rho_{11,1} & = & ( & 12& 0 & 0 & 0 & 0 & 0 &) \\ 
\boldsymbol\rho_{10,2} & = & ( & 66& 6 & 0 & 0 & 0 & 0 &) \\ 
\boldsymbol\rho_{9,3}  & = & ( &220& 0 & 4 & 0 & 0 & 0 &) \\
\boldsymbol\rho_{8,4}  & = & ( &495&15 & 0 & 3 & 0 & 0 &)\\
\boldsymbol\rho_{7,5}  & = & ( &792& 0 & 0 & 0 & 0 & 0 &)\\
\boldsymbol\rho_{6,6}  & = & ( &924&20 & 6 & 0 & 2 & 0 &).
\end{array}
\end{equation*}

The solutions (orbit number and length/symmetry), in order of
increasing symmetry, corresponding to orbit length $(12\;\;6\;\;4\;\;3\;\;2\;\;1)$
\begin{equation}\label{fixed-configuration12}
\begin{array}{lrrrrrrrrl}
\mathbf{A}_{12,0} & = & ( &  0& 0 & 0 & 0 & 0 & 1 &) \\ 
\mathbf{A}_{11,1} & = & ( &  1& 0 & 0 & 0 & 0 & 0 &) \\ 
\mathbf{A}_{10,2} & = & ( &  5& 1 & 0 & 0 & 0 & 0 &) \\ 
\mathbf{A}_{9,3}  & = & ( & 18& 0 & 1 & 0 & 0 & 0 &) \\
\mathbf{A}_{8,4}  & = & ( & 40& 2 & 0 & 1 & 0 & 0 &)\\
\mathbf{A}_{7,5}  & = & ( & 66& 0 & 0 & 0 & 0 & 0 &)\\
\mathbf{A}_{6,6}  & = & ( & 75& 3 & 1 & 0 & 1 & 0 &).
\end{array}
\end{equation}
The full H\"uckel cycles of Slater determinants are obtained by multiplying 
the orbit lengths appearing in~\eqnref{fixed-configuration12} by the 
size of the ring (13). Thus the Hubbard spectrum reads (units of $-2t$):
\begin{equation*}
\begin{array}{lccl}
M_S=6 (1)&:& \epsilon_{\lambda} = \cos\left(\frac{2\pi\lambda}{13} \right) & \lambda=0, \dots,\pm 5,\pm6  \\
M_S=5 (1)&:& \epsilon_{\lambda} = \cos\left(\frac{2\pi\lambda}{156}\right) & \lambda=0, \pm1,\dots, \pm 77, 78 \\
M_S=4 (5)&:& \epsilon_{\lambda} = \cos\left(\frac{2\pi\lambda}{156}\right) & \lambda=0, \pm1,\dots, \pm 77, 78 \\
M_S=4 (1)&:& \epsilon_{\lambda} = \cos\left(\frac{2\pi\lambda}{78}\right) & \lambda=0, \pm1,\dots, \pm 38, 39 \\
M_S=3 (18)&:& \epsilon_{\lambda} = \cos\left(\frac{2\pi\lambda}{156}\right) & \lambda=0, \pm1,\dots, \pm 77, 78 \\
M_S=3 (1)&:& \epsilon_{\lambda} = \cos\left(\frac{2\pi\lambda}{52}\right) & \lambda=0, \pm1,\dots, \pm 25, 26 \\
M_S=2 (40)&:& \epsilon_{\lambda} = \cos\left(\frac{2\pi\lambda}{156}\right) & \lambda=0, \pm1,\dots, \pm 77, 78 \\
M_S=2 (2)&:& \epsilon_{\lambda} = \cos\left(\frac{2\pi\lambda}{78}\right) & \lambda=0, \pm1,\dots, \pm 38, 39 \\
M_S=2 (1)&:& \epsilon_{\lambda} = \cos\left(\frac{2\pi\lambda}{39}\right) & \lambda=0, \pm1,\dots, \pm 18, \pm 19 \\
M_S=1 (66)&:& \epsilon_{\lambda} = \cos\left(\frac{2\pi\lambda}{156}\right) & \lambda=0, \pm1,\dots, \pm 77, 78 \\
M_S=0 (75)&:& \epsilon_{\lambda} = \cos\left(\frac{2\pi\lambda}{156}\right) & \lambda=0, \pm1,\dots, \pm 77, 78 \\
M_S=0 (3)&:& \epsilon_{\lambda} = \cos\left(\frac{2\pi\lambda}{78}\right) & \lambda=0, \pm1,\dots, \pm 38, 39 \\
M_S=0 (1)&:& \epsilon_{\lambda} = \cos\left(\frac{2\pi\lambda}{52}\right) & \lambda=0, \pm1,\dots, \pm 25, 26 \\
M_S=0 (1)&:& \epsilon_{\lambda} = \cos\left(\frac{2\pi\lambda}{26}\right) & \lambda=0, \pm1,\dots, \pm 12, 13 
\end{array}
\end{equation*}

Finally, an example of an odd-electron system, consisting of a ring with 22 metal centers
and 21 electrons. The parent symmetry of the necklace problem is now $\C_{21}$, with subgroups
$\{\C_1,\C_3,\C_7,\C_{21}\}$, giving rise to possible orbit lengths $\{21,7,3,1\}$.
The maximal spin of the system, corresponding to the configuration fully covering the 21 sites 
with a single `color' $X$, is equal to $M_S=21/2$. We thus have 22 possible spin-projection values,
and 11 unique configurations. For the purpose of illustrating the method we are going to sample
here only 5 spin states, namely $M_S=21/2$ ($X^{21}Y^0$), $M_S=15/2$ ($ X^{18}Y^{3}$), $M_S=7/2$ ($X^{14}Y^7$),
and the lowest spin state $M_S=1/2$ ($X^{11}Y^{10}$).
The inverse table of marks reads:
\begin{equation}\label{invmarks21}
\overline{\mathbf{m}} = \left( \begin{array}{rrrr}
\frac{1}{21}  &  0  &  0  &  0\\
-\frac{1}{21} &  \frac{1}{7} &  0  &  0\\
-\frac{1}{21}  &  0  &  \frac{1}{3}  &  0\\
\frac{1}{21}  &  -\frac{1}{7}  &  -\frac{1}{3}  &  1
\end{array}\right)
\end{equation}
and the $\boldsymbol\rho_{\theta}$ vectors for the four selected configurations are:
\begin{equation}\label{fixed-configuration21}
\begin{array}{rrrrrrrl}
\boldsymbol\rho_{21,0} & = & ( & 1 & 1 & 1 & 1 & ) \\ 
\boldsymbol\rho_{18,3} & = & ( & 1330 & 7 & 0 & 0 & ) \\ 
\boldsymbol\rho_{14,7} & = & ( & 116280& 0 & 3 & 0 & ) \\ 
\boldsymbol\rho_{11,10} & = & ( & 352716& 0 & 2 & 0 & ).
\end{array}
\end{equation}
leading to the solutions:
\begin{equation}\label{solutions21}
\begin{array}{rrrrrrrl}
\mathbf{A}_{21,0} & = & ( & 0 & 0 & 0 & 1 & ) \\ 
\mathbf{A}_{18,3} & = & ( & 63 & 1 & 0 & 0 & ) \\ 
\mathbf{A}_{14,7} & = & ( & 5537& 0 & 1 & 0 & ) \\ 
\mathbf{A}_{11,10} & = & ( & 16796 & 0 & 1 & 0 & ).
\end{array}
\end{equation}
 
\section{Generalization to any electron count: Eigenvalues for any $L<N$}\label{sec:generalN}

We present here the full solution of the eigenvalue spectrum for any $L<N$.
The unitary transformation employed here is due to Caspers and
Iske\cite{Caspers1989} and Kotrla.\cite{Kotrla1990} Our contribution is to use
permutation groups to determine the exact degeneracy of the solutions
determined in Refs.\ \onlinecite{Caspers1989,Kotrla1990}.

Consider first the case $M_S=L/2$, i.e., all electrons are spin-up:
$\bm{\sigma}=(\uparrow,\uparrow,\ldots,\uparrow)$. Then
$n_{i,\uparrow}n_{i,\downarrow}$ is necessarily zero and the Hubbard
Hamiltonian \eqref{hubbardHam} reduces, for \emph{any} value of $U$, to the
H\"uckel Hamiltonian of noninteracting electrons. Thus we immediately find that
the exact solutions for $M_S=L/2$ are given by \eqnref{EHuckel}, where every
H\"uckel molecular orbital $k$ can be occupied by at most one electron (because
of the Pauli principle). Notice that there is only one $\bm{\sigma}$ here
and the length of its orbit is $\omega_{\sigma}=1$. The number of $M_S=L/2$ states is
therefore equal to $\binom{N}{L}$.

This simple solution is of course not directly transferable to lower values of
$M_S$. However by a change of basis a connection with the maximum spin case can
be established.
Consider the cyclic permutation of the electron spins:
\begin{equation}
\hat{C}_L (\sigma_1,\sigma_2,\ldots,\sigma_L)=
(\sigma_L,\sigma_1,\ldots,\sigma_{L-1})
\end{equation}
Let us pick a certain subspace corresponding to a $\bm{\sigma}$-orbit, let $\omega_{\sigma}$ be the length of the orbit and let 
$\bm{\sigma}_0$ be a member of the orbit. $\bm{\sigma}_0$ can be thought of as
the representative spin configuration of that orbit. In fact, repeated application of
$\hat{C}_L$ cycles through all members of the orbit (leaving the occupation
vector $\vec{x}$ unchanged), so that $\hat{C}_L^{\omega_\sigma}=1$. It is not
difficult to see that $\hat{C}_L$ is a symmetry of our $U=\infty$ Hubbard
Hamiltonian.  Note that the determination of the number of spin configurations / necklaces $\bm{\sigma}$,
and the length $\omega_\sigma$ of their orbit under the effect of the group $\C_L$, represent the same combinatorial 
problem that has been solved by means of group theory in the previous paragraphs. 

Within a given $M_S$, the Hamiltonian is thus still block diagonal in $\bm{\sigma}$, where the dimension $n_\sigma$ of each block 
(only for the $L=N-1$ case representing the length of a H\"uckel annulene) is given by the 
product of the length of the orbit of spin configurations $\omega_\sigma$, times the number of possible 
orbital occupation vectors $\vec{x}$, which is $\binom{N}{L}$. We thus have a generalization of~\eqnref{HuckelOrbit}:
\[
n_\sigma = \binom{N}{L} \times \omega_\sigma.
\]

We proceed now to adapt the basis states of \eqnref{basis} to $\C_L$-symmetry:
\begin{equation}\label{symadap}
|\vec{x},\bm{\sigma}_0,\kappa\rangle=\frac{1}{\sqrt{\omega_\sigma}}
\sum_{n=0}^{\omega_\sigma-1} 
e^{\frac{i 2\pi\kappa}{\omega_\sigma}n} |\vec{x},\hat{C}_L^n \bm{\sigma}_0\rangle,
\qquad \kappa=0,1,\ldots,\omega_\sigma-1
\end{equation}
which causes a further division of the $\bm{\sigma}$-subspace into $\omega_\sigma$ sub-subspaces, denoted
$\{\bm{\sigma}_0,\kappa\}$. Note that each space $\{\bm{\sigma}_0,\kappa\}$
consists of $\binom{N}{L}$ states, corresponding to the possible occupation
vectors $\vec{x}$. Now using \eqnref{symadap} it is not difficult to
show\cite{Caspers1989,Kotrla1990} that the matrix of $H_t$ in this space
is the same as the matrix of a \emph{modified} $H_t'$ in the space of
$M_S=L/2$ (whose basis states we denote here simply by $|\vec{x}\rangle$):
\begin{equation}\label{mapping}
\langle \vec{x}_i,\bm{\sigma}_0,\kappa|H_t| 
\vec{x}_j,\bm{\sigma}_0,\kappa\rangle = 
\langle \vec{x}_i | H_t' | \vec{x}_j \rangle,
\end{equation}
The modified $H'_t$ is obtained from $H_t$ by adding a phase to the hopping integral
between site N and 1:
\begin{equation}
\langle N|H'_t|1\rangle = e^\frac{i 2\pi\kappa}{\omega_\sigma} t = 
\langle 1|H'_t|N\rangle^*
\end{equation}
\eqnref{mapping} thus establishes a correspondence between the subspace
$\{\bm{\sigma}_0,\kappa\}$ of the $U=\infty$ Hubbard ring and a fictitious
system of all spin-up (or, equivalently, spinless), noninteracting electrons
on a H\"uckel ring described by $H'_t$. The solutions of the latter are easy to
obtain. The H\"uckel molecular orbitals and energies are given by
\begin{align}
a_\lambda^\dagger&=\frac{1}{\sqrt{N}}\sum_{n=1}^N
e^{i\frac{2\pi}{N}(\lambda+\frac{\kappa}{\omega_\sigma})n} \,c_n^\dagger,\\
\varepsilon_\lambda&=
2t\cos\left[\frac{2\pi}{N}\left(\lambda+\frac{\kappa}{\omega_\sigma}\right)\right],
\end{align}
where $\lambda$ can take the values given by Eqs.\ \eqref{kLodd} and \eqref{kLeven}.
Occupying the H\"uckel orbitals with $L$ electrons gives the total energy
\begin{equation}
E=\sum_{j=1}^L
2t\cos\left[\frac{2\pi}{N}\left(\lambda_j+\frac{\kappa}{\omega_\sigma}\right)\right],
\end{equation}
where no two electrons can occupy the same orbital. This concludes the complete
diagonalization of the $U=\infty$ Hubbard ring.

\begin{acknowledgements} 
A.S. and W.V.d.H would like to thank A. Dao for useful discussions. 
A.S. acknowledges support from the Selby Research Award, and the Early Career Researcher 
grant scheme from the University of Melbourne.
\end{acknowledgements}

\appendix*

\section*{Appendix A: generation of table of marks for cyclic groups}

The full set of cosets (or coset decomposition) of the cyclic group $\C_L$ generated
by the subgroup $\C_k$, for any $k$ that is divisor of $L$,
can be written as:
\begin{equation}\label{coset}
\C_L/\C_k = \{\C_k, \C_k \hat{C}_L^{1}, \C_k \hat{C}_L^{2}\dots, \C_k \hat{C}_L^{\frac{L}{k}-1} \}
\end{equation}
Also, the cycle-structure of the permutations belonging to the regular representation 
of the $\C_L$ group can be easily determined.
The $\C_L$ group contains $L$ rotations, $\hat{C}_L^{0}\equiv\hat{1}$, $\hat{C}_L^{1}$, $\hat{C}_L^{2}$\dots
$\hat{C}_L^{k}$\dots$\hat{C}_L^{L-1}$.
The permutation associated to the identity operator $\hat{C}_L^{0}\equiv\hat{1}$ in a regular representation is simply 
decomposed into $L$ 1-cycles: $(1)(2)\dots(L)$, and the cycle-structure of the permutation associated 
to a $\hat{C}_L^{1}$ rotation is always a single $L$-cycle $(1,2,3\dots,L)$. 
For a general $k < L$, the cycle-structure of the permutation associated to the rotation $\hat{C}_L^{k}$ 
in a regular representation
consists of $m$ $(L/m)$-cycles, i.e.\ of $m$ cycles, all of the same size $L/m$, where $m=\mathrm{gcd}(L,k)$.
Since the regular representation is a faithful representation (i.e.\ each permutation corresponds
to one and only one of the $L$ rotations of the $\C_L$ group), and each permutation of a regular 
representation is decomposed in cycles of equal size, it follows that only the identity operator 
in a regular representation contains 1-cycles ($L$ of them).

To build a table of marks it is important to identify in each coset representation $\C_L(/\C_k)$,
those permutations that contain 1-cycles in their cycle-decomposition, as 1-cycles correspond to
fixed-cosets in the coset representation $\C_L(/\C_k)$ under the action of some subgroup. 
We can thus proceed as follows. 

First we build each coset representation $\C_L(/\C_k)$
by acting with the group $\C_L$ on the set of cosets~\eqnref{coset}, as described in~\eqnref{permutation}.
This gives rise, for $k>1$, to a permutation representation that is clearly not faithful, 
meaning that for $k>1$ the same permutation is repeated more than once in $\C_L(/\C_k)$.  
It is in fact easy to show that the coset representation $\C_L(/\C_k)$ is equivalent 
to $k$ copies of the regular representation of the cyclic group $\C_{L/k}$. Thus, the identity operator
of the regular representation of $\C_{L/k}$ (whose domain consists of $L/k$ points, i.e. the $L/k$
right cosets of the set of cosets $\C_L/\C_k$ in~\eqnref{coset}) consists of $(L/k)$ 1-cycles, and 
is repeated $k$-times within the coset representation $\C_L(/\C_k)$. The $k$ identity operators 
(representing the operators $\hat{C}_L^{q\frac{L}{k}}$, $q=0,1,\dots,k-1$, of the parent group $\C_L$), 
are the only permutations in $\C_L(/\C_k)$ that contain 1-cycles at all.

Next, now that we know the detailed structure of all coset representations $\C_L(/\C_k)$, we proceed to build
the table of marks $m_{kj}$ by acting with all operations of the subgroup $\C_j$, for all $j=1,\dots,s$, 
on the set of cosets $\C_L/\C_k$~\eqnref{coset}, and by counting how many remain fixed under the action of $\C_j$.
This is equivalent to inspecting the coset representation $\C_L(/\C_k)$, and counting how many
1-cycles are shared between the representations in $\C_L(/\C_k)$ of {\it all} operations of $\C_j$. 
The 1-cycles in the cycle-structure of a permutation associated to one operation of $\C_j$,
correspond in fact to cosets that are fixed under the action of that particular operation of  $\C_j$.
If all operations of the subgroup $\C_j$ correspond to permutations in $\C_L(/\C_k)$ sharing
a number $n$ of 1-cycles, then $n$ is the mark $m_{kj}$. 

Since we have established that $\C_L(/\C_k)$ contains only $k$ permutations with 1-cycles,
in fact made of $L/k$ 1-cycles, corresponding to the $\C_L$-operations $\hat{C}_L^{q\frac{L}{k}}$, $q=0,1,\dots,k-1$, 
it follows that only if $j$ is a divisor of $k$ then all operations of $\C_j$ correspond to a subset of 
the $\hat{C}_L^{q\frac{L}{k}}$ operations represented in terms of $L/k$ 1-cycles, thus the mark $m_{kj}$ is non-zero 
and equal to $L/k$. 
Thus we can write an analytical expression for the table of marks of any cyclic group $\C_L(/\C_k)$ as:
\begin{equation}\label{marktable}
m_{kj}=\begin{cases}
  L/k & \text{if $j\mid k$}, \\
  0 & \text{otherwise}.
 \end{cases}
\end{equation}

The marks of $\C_L$ for $L=6$ and $L=12$ generated via~\eqnref{marktable} are
reported in Tables \ref{tbl:C6} and \ref{tbl:C12}.

\begin{table}
\caption{Table of marks for cyclic group $\C_6$}
\label{tbl:C6}
\begin{tabular}{l|rrrr}
  $\C_6$         & $\C_1$ & $\C_2$ & $\C_3$ & $\C_6$ \\ \hline
  $\C_6(/\C_1)$  & 6      & 0      & 0      & 0       \\
  $\C_6(/\C_2)$  & 3      & 3      & 0      & 0       \\
  $\C_6(/\C_3)$  & 2      & 0      & 2      & 0       \\
  $\C_6(/\C_6)$  & 1      & 1      & 1      & 1       \\
\end{tabular}
\end{table}

\begin{table}
\caption{Table of marks for cyclic group $\C_{12}$}
\label{tbl:C12}
\begin{tabular}{l|rrrrrr}
  $\C_{12}$            & $\C_1$ & $\C_2$ & $\C_3$ & $\C_4$ & $\C_6$ & $\C_{12}$\\ \hline
  $\C_{12}(/\C_1)$     & 12     & 0      & 0      & 0      & 0      & 0        \\
  $\C_{12}(/\C_2)$     & 6      & 6      & 0      & 0      & 0      & 0        \\
  $\C_{12}(/\C_3)$     & 4      & 0      & 4      & 0      & 0      & 0        \\
  $\C_{12}(/\C_4)$     & 3      & 3      & 0      & 3      & 0      & 0        \\
  $\C_{12}(/\C_6)$     & 2      & 2      & 2      & 0      & 2      & 0        \\
  $\C_{12}(/\C_{12})$  & 1      & 1      & 1      & 1      & 1      & 1        \\
\end{tabular}
\end{table}

\section*{Appendix B: subduced coset representations and orbits of
configurations}

\subsection*{B.1. Subduced coset representations and suborbits }\label{subduction}

A coset representation $\G(/\G_i)$ associated to a subgroup $\G_i$ is  obtained by acting with
the parent group $\G$ on the coset decomposition of $\G$ generated by $\G_i$. The parent group $\G$
acting on a domain $\Delta$ partitions it into  a number of orbits $\Delta_i$ according to ~\eqnref{crdecomposition},
where each orbit $\Delta_i$ correspond to a coset representation $\G(/\G_i)$.
If we now act with a subgroup $\G_j$ on each orbit $\Delta_i$, we obtain a partition of each orbit $\Delta_i$
into suborbits $\Delta_{ij,k}$, where $k$ runs over the subgroups of the subgroup $\G_j$.

This further partition of the domain $\Delta$ into suborbits can also be described in terms of subduced 
coset representations generated by $\G_j$. 
The subduced coset representation $\G(/\G_i)\downarrow\G_j$ is obtained by acting on the same coset 
decomposition of $\G$ generated by $\G_i$, with the subgroup $\G_j$.  Whereas $\G(/\G_i)$ is 
always a transitive representation, $\G(/\G_i)\downarrow\G_j$ clearly is not, as the domain represented 
by the right cosets generated by $\G_i$, a single orbit under the action of $\G$, will be
partitioned into sub-orbits under the action of $\G_j$. 

Given the set of $m_j$ subgroups of $\G_j$, $\lambda=\{\G_1^{(j)},\G_2^{(j)},\dots\G_k^{(j)},\dots \G_{m_j}^{(j)}  \}$
it is thus possible to reduce the intransitive subduced coset representation $\G(/\G_i)\downarrow\G_j$ 
to a sum of transitive coset representations generated by the subgroups $\lambda$,
by applying~\eqnref{crdecomposition}:
\begin{equation}\label{subducedcr}
\G(/\G_i)\downarrow\G_j = \sum_{k=1}^{m_j}\beta_k^{(ij)}\G_j(/\G_k^{(j)})
\end{equation}
where $\beta_k^{(ij)}$ are multiplicities, i.e.\ number of sub-orbits ruled by the CR $\G_j(/\G_k)$,
of size $d_{jk}=|\G_j|/|\G_k^{(j)}|$,
subduced from a single orbit associated to the CR $\G(/\G_i)$ under the action of $\G_j$.

We can also easily find a reduction formula which allows to compute straight away the multiplicities $\beta_k^{(ij)}$:
\begin{equation}\label{reductiosub0}
\mu_q^{(j)} = \sum_{k=1}^{m_j} \beta_k^{(ij)} m^{(j)}_{kq}, \;\; q=1, 2,\dots m_j,
\end{equation}
where the matrices $m^{(j)}_{kq}$ correspond to the table of marks of the subgroup $\G_j$.
Thus, the quantities appearing in~\eqnref{reductiosub0} can be easily precomputed once and for all.
In fact, the number of fixed points (cosets) $\mu_q^{(j)}$ in $\G(/\G_i)\downarrow\G_j$ under the action of
$\G_q$ can be simply retrieved from the table of marks $m_{iq}$ for $\G$, selecting only those columns $q$
that correspond to the subgroups of $\G_j$ (with some complication arising if the parent group and the subgroup do not
share the same structure of conjugacy classes). The resulting rectangular matrix with $s$ rows (number of subgroups 
of $\G$) and $m_j$ columns (number of subgroups of $\G_j$), obtained from the square matrix
table of marks for $\G$, is known as {\it subduced mark table}, with $M^{(j)}_{iq}\equiv\mu_q^{(j)}, i=1,\dots,s$.
Thus we can calculate the multiplicities $\beta_k^{(ij)}$ by inverting~\eqnref{reductiosub0}:
\begin{equation}\label{reductiosub}
\beta_k^{(ij)} = \sum_{q=1}^{m_j} M^{(j)}_{iq} \overline{m}^{(j)}_{qk}.
\end{equation}
where $\overline{m}^{(j)}_{qk}$ is the inverse of the square matrix $m^{(j)}_{kq}$ (inverse of table of marks
for subgroup $\G_j$).
 
\subsection*{B.2. Orbits of configurations}
A strategy to compute $\rho_{\theta j}$ can be devised by noticing which conditions a given 
function/configuration $f^{\theta}_k$ has to fulfill in order to be constant under 
the action of all operations of the subgroup $\G_j$.
By definition, for all $g\in\G_j$ and all $\delta\in\Delta$, an invariant configuration $f^{\theta}$ must 
obey $f^{\theta}(p_g(\delta))=f^{\theta}(\delta)$.  The operations $p_g(\delta)$ in turn,  $\forall g\in\G_j$, 
partition each orbit $\Delta_i$ (generated by the parent group $\G$ on the domain $\Delta)$ into $\G_j$-suborbits.
The problem of determining the number and size of the suborbits generated by the action of the subgroup $\G_j$ 
on a given orbit $\Delta_i$ ruled by the coset representation $\G(/\G_i)$ has been solved in subsection~\ref{subduction},
namely, by reducing the intransitive subduced coset representation  $\G(/\G_i)\downarrow\G_j$ into
transitive coset representations of the subgroup $\G_j$, via~\eqnref{subducedcr}, \eqnref{reductiosub0} and~\eqnref{reductiosub}
With this information (number and length of $\G_j$-subduced orbits), we can now build explicitly a generating function 
counting the number of fixed configurations $f^{\theta}_k$ under the action of the subgroup $\G_j$,
in the form of a symmetry-adapted polynomial in the 'colors' $X$ and $Y$, where the number of 
configurations for a given partition (spin) $\theta = \{n_1,n_2\}$  corresponds to 
the coefficient of the monomial (or weight) $X^{n_1} Y^{n_2}$.

Thus according to~\eqnref{subducedcr},~\eqnref{reductiosub0}, and~\eqnref{reductiosub}, the action of the subgroup $\G_j$
on each orbit $\Delta_i$ is described by a subduced representation $\G(/\G_i)\downarrow\G_j$, which 
partitions the orbit $\Delta_i$ into $\beta_{k}^{(ij)}$ suborbits of length $d_{jk}=|\G_j|/|\G_k^{(j)}|$
(where $\G_k^{(j)}$ are subgroups of $\G_j$, for $k=1,\dots,m_j$), each suborbits corresponding 
to the coset representation $\G_j(/\G_k^{(j)})$.  In order for $f^{\theta}_k$ to be constant, each 
suborbit of length $d_{jk}$ has to be decorated by beads of the same color, leading to either $d_{jk}$
beads of color $X$, or $d_{jk}$ beads of color $Y$ in this particular case. 
Hence it is straightforward to write down the generating function 
for each suborbit (also known in P{\'o}lya-Redfield theory as the {\it figure inventory}) as:
\begin{equation}\label{figureinventory}
s_{d_{jk}}=X^{d_{jk}}+Y^{d_{jk}}.
\end{equation}

Next, we need to extend the definition of the generating function~\eqnref{figureinventory} so to take 
into account all the $\beta_k^{(ij)}$ suborbits of symmetry $\G_k$ (as $\left(s_{d_{jk}}\right)^{\beta_k^{(ij)}}$), 
and by multiplying together all the resulting inventories for all possible subgroups of $\G_j$.
It can be readily seen that this process leads to the definition of the Fujita's {\it Unit Subduced 
Cycle Indices} (USCIs) $Z(\G(/\G_i)\downarrow\G_j; s_{d_{jk}})$ as~\cite{Fujita1989}
\begin{equation}\label{usci}
Z(\G(/\G_i)\downarrow\G_j; s_{d_{jk}}) = \prod_{k=1}^{m_j} \left( s_{d_{jk}}\right)^{\beta_{k}^{(ij)}}
\end{equation}
which in this particular case of two-colors only reduces to:
\begin{equation}\label{uscitwocolors}
Z(\G(/\G_i)\downarrow\G_j; s_{d_{jk}}) = \prod_{k=1}^{m_j} \left(X^{d_{jk}}+Y^{d_{jk}}\right)^{\beta_{k}^{(ij)}}.
\end{equation}

Finally, by taking into account all original orbits $\Delta_i$ into which the domain $\Delta$ is partitioned
by the action of $\G$, including their multiplicities $\alpha_i$ given by~\eqnref{crdecomposition},
we obtain the final generating function or {\it Unit Cycle Index} $Z(\G_j;s_{d_{jk}}^{(\alpha)})$  (UCI)
associated to the symmetry $\G_j$ as
\begin{equation}\label{uci}
Z(\G_j; s_{d_{jk}}^{(\alpha)}) = \prod_{i=1}^{s}\prod_{\alpha=1}^{\alpha_i} Z(\G(/\G_i)\downarrow\G_j; s_{d_{jk}}^{(\alpha)})
\end{equation}
where the superscript $\alpha$ in  $s_{d_{jk}}^{(\alpha)}$ indicates the possibility of assigning different 
figure-inventories~\eqnref{figureinventory} to different orbits of the original domain $\Delta$ 
(useful e.g. to assign different chemical valency to atoms belonging to different orbits in chemical enumeration).
Clearly, since all configurations of symmetry $\G_j$ are invariant under the action of the group $\G_j$,
the UCI must also equal the sum over all possible partitions $\theta$ of the weight of that particular
partition ($X^{n_1} Y^{n_2}$) times the number of configurations $f^{\theta}$ that are left invariant under 
the action of $\G_j$, i.e. the mark $\rho_{\theta j}$ appearing in the reduction 
formula~\eqnref{reductio2}.  This observation leads to a practical recipe to build a symmetry-adapted polynomial, 
where, for each given symmetry $\G_j$, the coefficients of the weights ($X^{n_1} Y^{n_2}$) 
are the marks $\rho_{\theta j}$: 
\begin{equation}\label{generatingfunction}
\sum_{\substack{ \theta \\ \{n_1,n_2\}}} 
\rho_{\theta j} X^{n_1}Y^{n_2} = 
\prod_{i=1}^{s}\prod_{\alpha=1}^{\alpha_i}\prod_{k=1}^{m_j} (X^{d_{jk}}+Y^{d_{jk}})^{\beta_{k}^{(ij)}}
\end{equation}

Since $s,  m_j, d_{jk}$ are known and universal,  
and $\alpha_i$ and $\beta_k^{(ij)}$ can be computed from the table of marks using equations~\eqnref{reductio1}
and~\eqnref{reductiosub}, it follows that~\eqnref{generatingfunction} provides a clear strategy to compute
the marks $\rho_{\theta j}$. Substitution of the marks $\rho_{\theta j}$ into the system of linear 
equations~\eqnref{reductio2} leads to the determination of the multiplicities $A_{\theta j}$, thus 
to the full solution of the Hubbard problem for any ring size $N$ and spin projection $M_S$.

\subsection*{B.3. Hubbard-H\"uckel rings: orbits of configurations in cyclic groups}

Things are further simplified when we try to apply the two fundamental equations~\eqnref{generatingfunction}
and~\eqnref{reductio2} to the problem of counting inequivalent $L$-necklaces {\it of a given symmetry} 
arising from 2-colors decoration of an $L$-necklace, whose symmetry is that 
of the cyclic group $\C_L$. First of all, the $L$-ring is a single orbit of size $L$ of the 
group $\C_L$, it is thus ruled by the regular coset representation $\C_L(/\C_1)$.
For this case the calculation of $\beta_{k}^{(ij)}$, for $i=1$, is greatly simplified.
In fact, by ordering subgroups of $\C_n$ in increasing group-order ($C_1$ is thus the first and $\C_L$ the last),
the table of marks for $\C_L$ becomes a lower triangular matrix~\cite{Fujita1989} (see also examples in
the appendix).  Thus the only non-zero element of the first row is the first element $m_{11}$, clearly equal to $|\C_L/\C_1|=L$
(the identity subgroup $\C_1$ leaves invariant all $L$ cosets of $\C_L(/\C_1)$, thus the mark of $\C_1$ in $\C_L(/\C_1)$ is $L$).  
This holds also for the subduced marks table $M^{(j)}$, since $\C_1$ is subgroup to all subgroups of $\C_L$.
Hence equation~\eqnref{reductiosub} reduces to the calculation of only the first row of  $\overline{m}^{(j)}_{qk}$
\[
\beta_{k}^{(1,j)}= L \overline{m}^{(j)}_{1k}.
\]
Furthermore, since the table of marks for the subgroup $\C_j$ has only the first element that is non-zero
and equal to $|\C_j|=j$, it follows that also its inverse has only the first element that is non-zero,
and equal to $1/j$.  Thus  $\beta_{k}^{(1j)} $ is non-zero only if $k=1$, leading to the only possibility
\[
\beta_1^{(1j)}=\frac{L}{j},
\] 
and the simple reduction of the single orbit $\C_L/\C_1$
to $L/j$ suborbits $\C_j/\C_1$ of maximal length $d_{j1}=j$, a fact expressed in terms of equation~\eqnref{subducedcr} as:
\[
\C_L(/\C_1)\downarrow\C_j = \frac{L}{j}\C_j(/\C_1).
\] 

It follows immediately that the equation for the number of fixed-configurations $\rho_{\theta j}$
of a given symmetry $\G_j$~\eqnref{generatingfunction} found in the previous subsection simplifies to:
\begin{equation}\label{cyclicgenerating}
\sum_{\theta} \rho_{\theta j} X^{n_1} Y^{n_2}= (X^{j}+Y^{j})^{\frac{L}{j}}.
\end{equation}
Finally, binomial expansion of the rhs of~\eqnref{cyclicgenerating} gives:
\begin{equation}\label{binomialexpansion}
(X^{j}+Y^{j})^{\frac{L}{j}} = \sum_{k=0}^{L/j} 
\binom{L/j}{k} X^{L-kj}Y^{kj}.
\end{equation}
Thus a general formula for the mark of symmetry $j$ ($\rho_{\theta j}$) for each 
partition $\theta=\{n_1,n_2\}$, can be obtained by equating the rhs of~\eqnref{binomialexpansion}
to the lhs of~\eqnref{cyclicgenerating}, leading to:
\begin{equation}\label{fpv}
\rho_{\theta j} = \binom{L/j}{n_1/j}.
\end{equation}
Clearly, not all partitions $\theta$ will be allowed in a given cyclic subgroup $\C_j$, but only
those partitions for which $j$, the order of the cyclic subgroup, is a divisor of $n_1$ and $n_2$.
Thus the multiplicities $A_{\theta j}$ (i.e. number of orbits of length $L/j$) for a fixed configuration
(spin) $\theta$ can be determined simultaneously for all symmetries $j$ (all divisors of $L$)
by a simple vector-matrix multiplication, by computing~\eqnref{reductio2} for a given $\theta$,
and for all symmetries $j=1,\dots,s$ ($s$ is the number of divisors of $L$), and by inverting the resulting 
matrix equation. If we collect the multiplicities $A_{\theta j}$ for a given configuration $\theta$ and for 
all symmetries in the $1\times s$ row-matrix $\mathbf{A}_{\theta}$, 
the marks~\eqnref{fpv} for all symmetries and given $\theta$ in the $1\times s$  row matrix  
$\boldsymbol\rho_{\theta}$, and the inverse table of marks $\overline{m}_{ji}$ for the
group $\C_L$ in the $s\times s$ square matrix $\overline{\mathbf{m}}$, we obtain the general 
solution to the symmetry adapted two-color necklace problem as :
\begin{equation}\label{solutionvector}
\mathbf{A}_{\theta} = \boldsymbol\rho_{\theta} \overline{\mathbf{m}}
\end{equation}


\begin{thebibliography}{12}

\bibitem{Mott1968} N. F. Mott, Rev. Mod. Phys. \textbf{40}, 677 (1968).
\bibitem{Gatteschi2006} D.~Gatteschi, R. Sessoli, and J. Villain,
\textit{Molecular Nanomagnets} (Oxford University Press, Oxford, 2006).
\bibitem{EsslerBook2005} F.~H.~Essler, H.~Frahm, F.~G\"ohmann, A.~Kl\"umper and V.~E.~Korepin,
\textit{The One-Dimensional Hubbard Model} (Cambridge University Press, Cambridge, 2005).
\bibitem{Tasaki1998a} H.~Tasaki, Prog. Theor. Phys. {\bf 99}, 489 (1998).
\bibitem{Tasaki1998b} H.~Tasaki, J. Phys.: Condens. Matter {\bf 10}, 4353 (1998).
\bibitem{Caspers1989} W.~J.~Caspers and P.~L.~Iske, Physica A {\bf 157}, 1033 (1989).
\bibitem{Kotrla1990} M.~Kotrla,  Physics Letters A {\bf 145}, 33 (1990).
\bibitem{BruusFlensberg2004} H.~Bruus and K.~Flensberg, 
\textit{Many-body Quantum Theory in Condensed Matter Physics} (Oxford University Press, Oxford, 2004). See Chapter 10.
\bibitem{ElsteTimm2005} F.~Elste and C.~Timm, Phys. Rev. B {\bf 71}, 155403 (2005).
\bibitem{SonciniJACS2010} A.~Soncini, T.~Mallah and L.~F.~Chibotaru, J. Am. Chem. Soc. {\bf 132}, 8106 (2010).
\bibitem{SonciniPRB2010} A.~Soncini and L.~F.~Chibotaru, Phys. Rev. B {\bf 81}, 132403 (2010). 
\bibitem{SteinerFowlerJPCA2001} E.~Steiner and P.~W.~Fowler, J. Phys. Chem. A {\bf 105}, 9553 (2001).
\bibitem{SteinerFowlerChemComm2001} E.~Steiner and P.~W.~Fowler, Chem. Comm., 2220 (2001).
\bibitem{Burnside1911} W.~Burnside, \textit{Theory of groups of finite order},
2nd ed.\ (Cambridge University Press, Cambridge, 1911).
\bibitem{Fujita1989} S.~Fujita, Theor. Chim. Acta {\bf 76}, 247 (1989).
\bibitem{FowlerQuinn1986} P.~W.~Fowler and C.~M.~Quinn, Theor. Chim. Acta  {\bf 70}, 333 (1986).
\bibitem{CeulemansFowler1991} A.~Ceulemans and P.~W.~Fowler, Nature {\bf 353}, 52 (1991).
\bibitem{Harary1973} F.~Harari and E.~M.~Palmer, \textit{Graphical
Enumeration} (Academic Press, New York, 1973).
\bibitem{Hardy2008} G.~H.~Hardy and E.~M.~Wright, \textit{An introduction to
the theory of numbers}, 6th ed.\ (Oxford University Press, Oxford, 2008).

\end{thebibliography}
\end{document}